\documentclass[iop,twocolappendix,numberedappendix,appendixfloats]{emulateapj}
\newcommand{\Msun}{\,M_{\odot}}

\newcommand\ionn[2]{#1$\;${\scshape{#2}}}
\def\kms{\ {\rm km\, s}^{-1}}
\def\teff{T_{\rm eff}}
\def\logg{{\rm log}\,g}

\def\alf{\texttt{alf}}

\shortauthors{CONROY ET AL.}

\shorttitle{Updated Stellar Population Models for Old Stellar Systems}

\begin{document}

\title{Metal-rich, Metal-poor: Updated Stellar Population Models for Old Stellar Systems}

\author{Charlie Conroy\altaffilmark{1},
  Alexa Villaume\altaffilmark{2},
  Pieter G. van Dokkum\altaffilmark{3},
Karin Lind\altaffilmark{4}}

\altaffiltext{1}{Department of Astronomy, Harvard University,
  Cambridge, MA, 02138, USA}
\altaffiltext{2}{Department of Astronomy and Astrophysics, University
  of California, Santa Cruz, CA 95064, USA}
\altaffiltext{3}{Department of Astronomy, Yale University, New Haven,
  CT, 06511, USA}
\altaffiltext{4}{Max Planck Institute for Astronomy, Königstuhl 17, D-69117 Heidel- berg, Germany}

\slugcomment{Submitted to ApJ}

\begin{abstract}

  We present updated stellar population models appropriate for old
  ages ($>1$ Gyr) and covering a wide range in metallicities
  ($-1.5\lesssim$[Fe/H]$\lesssim0.3$).  These models predict the full
  spectral variation associated with individual element abundance
  variation as a function of metallicity and age.  The models span the
  optical-NIR wavelength range ($0.37-2.4\mu m$), include a range of
  initial mass functions (IMFs) and contain the flexibility to vary 18
  individual elements including C, N, O, Mg, Si, Ca, Ti, and Fe.  To
  test the fidelity of the models we fit them to integrated light
  optical spectra of 41 Galactic globular clusters (GCs).  The value
  of testing models against GCs is that their ages, metallicities, and
  detailed abundance patterns have been derived from the HR diagram in
  combination with high resolution spectroscopy of individual stars.
  We determine stellar population parameters from fits to all
  wavelengths simultaneously (``full spectrum fitting"), and
  demonstrate explicitly with mock tests that this approach produces
  smaller uncertainties at fixed S/N ratio than fitting a standard set
  of 14 line indices.  Comparison of our integrated-light results to
  literature values reveals good agreement in metallicity, [Fe/H].
  When restricting to GCs without prominent blue horizontal branch
  populations we also find good agreement with literature values for
  ages, [Mg/Fe], [Si/Fe], and [Ti/Fe].

\end{abstract}



\section{Introduction}
\label{s:intro}

The physical properties of stellar populations provide key insight
into their formation and assembly histories.  Stellar ages enable
direct insight into when populations formed; metallicities provide
clues to chemical enrichment, stellar feedback, and the baryon cycle
into and out of galaxies; detailed elemental abundance patterns can be
used to estimate star formation timescales and can be used as `tags'
to trace populations across epochs \citep[e.g.,][]{Choi14}; and the
measurement of the stellar initial mass function across the galaxy
population provides a new view of the star formation process in
extreme environments.  

For resolved stellar populations, these parameters can be measured
relatively straightforwardly, either from CMDs, star counts, or high
resolution spectra of individual stars.  Many of these parameters are
difficult to measure from the integrated light of composite stellar
populations (including star clusters and galaxies), where the light
from $\sim10^3-10^{12}$ stars are combined into a single spectrum.
Nonetheless, substantial progress has been made over the past several
decades.  In the field of modeling continuum (absorption line)
spectra, the majority of models have been built to interpret spectral
indices, i.e., equivalent widths of a handful ($3-20$) of features,
mostly in the optical window \citep[e.g.,][]{Worthey94, Trager00,
  Thomas05, Schiavon07, Thomas11}.  The focus on indices was partly
historical in that early spectral observations of early-type galaxies
were of relatively low spectral resolution (FWHM $\sim10$\AA), and
partly an attempt to focus attention on the spectral features that
showed the greatest sensitivity to the main parameters of interest,
e.g., age, metallicity, and $\alpha$-enhancement \citep{Worthey94}.

More recent modeling efforts have highlighted the advantages of
modeling the full absorption line spectrum in order to derive detailed
abundance patterns \citep[e.g.,][]{Conroy12a, Conroy14a}. At a
fundamental level, the information content is higher in the native
spectrum than in derived equivalent widths measured over
$\approx10-50$\AA\, windows.  Many parameters, such as the IMF
\citep{Conroy12a} and trace elements \citep{Conroy13a}, have subtle
effects on the spectrum, and their effects are much easier to identify
in the spectra than in indices.  As an example, the effect of
increasing the low-mass IMF results in stronger \ionn{Na}{i}
absorption at $0.82\mu m$.  However, this feature is strongly blended
with TiO absorption, which has a slightly bluer maximum absorption
depth.  Increasing the contribution of low-mass stars therefore
induces not only stronger absorption at $\approx 0.82\mu m$ but also a
redward shift in the blended absorption feature \citep{vanDokkum12}.
This important signature is lost when measuring this feature with an
index.  In addition, for many cases the information is contained in
faint features spread over a wide wavelength range, rather than in one
or two localized features.

One of the key ingredients in developing high quality stellar
population models is the input stellar spectral libraries.  The
standard approach is to use empirical libraries as the backbone of the
model and to rely on theoretical spectra to ``correct'' the empirical
spectra to arbitrary elemental abundance patterns (these are often
referred to as ``response functions'').  The current state-of-the-art
in the optical wavelength range is the MILES spectral library
\citep{Sanchez-Blazquez06} which has wide coverage in $\teff$,
$\logg$, and metallicity, and is used in many stellar population
models \citep[e.g.,][]{Vazdekis10, Thomas11, Conroy12a}.

Stellar libraries in the NIR ($>0.8\mu m$) are only now becoming
comparable to the optical in terms of quality and quantity.  A major
advance was the IRTF Cool Star Library \citep{Rayner09} of 210 stars
which has since been incorporated into several stellar population
models \citep{Conroy12a, MenesesGoytia15b, Rock16}.  The major
shortcoming of that library was the narrow coverage in metallicity.
To overcome this limitation we acquired IRTF spectra of 284 stars in
the MILES library and presented this new, Extended IRTF Library in
\citet{Villaume17a}.  These stars were chosen to span a wide range in
$\teff$, $\logg$, and metallicity in order to develop new stellar
population models covering a wide metallicity and wavelength range.
These new models are the focus of this paper.

\begin{figure*}[!t]
\center
\includegraphics[width=0.9\textwidth]{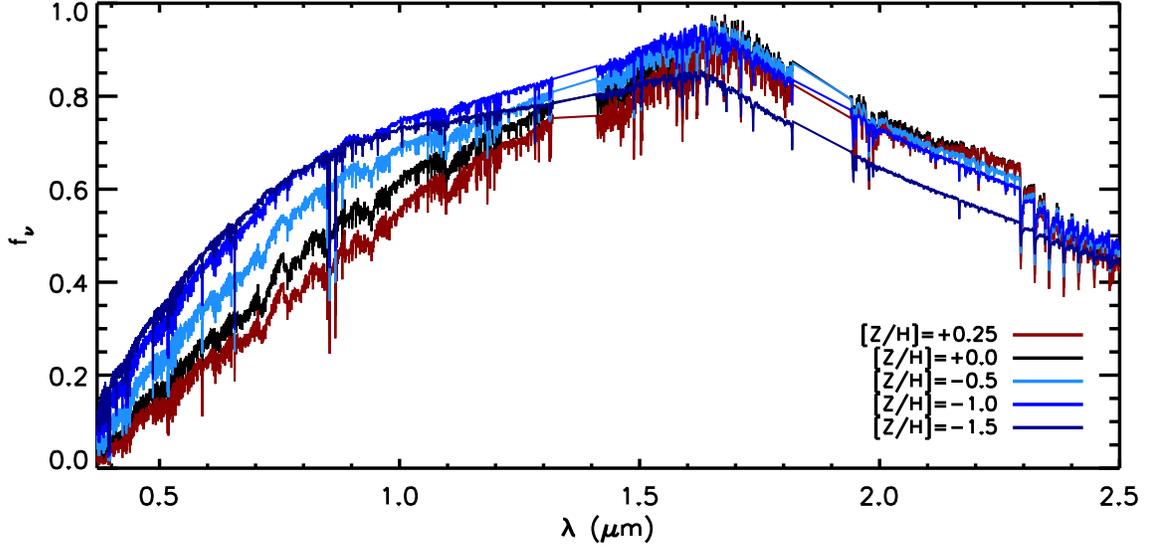}
\vspace{0.1cm}
\caption{Empirical SSPs as a function of metallicity at 13 Gyr.}
\label{fig:sspsum}
\end{figure*}


\section{Models}
\label{s:models}

In this section we describe the construction of the empirical simple
stellar populations (SSPs; Section \ref{s:emp}) and the theoretical
response functions (Section \ref{s:rfn}).  Where relevant, we
highlight updates with respect to our previous generation of models
presented in \citet{Conroy12a, Conroy14a, Choi14}.  The single most
important improvement with respect to our previous models is the
much larger metallicity coverage in the new models.  The older models
were valid only for modest deviations from solar metallicity while
these new models span nearly the entire relevant metallicity range for
globular clusters and galaxies ($-1.5\lesssim$[Fe/H]$\lesssim0.3$).

\begin{deluxetable}{rccc}
\tablecaption{Adopted Abundance Pattern of Library Stars}
\tablehead{ \colhead{[Fe/H]} & \colhead{[O/Fe]} & \colhead{[Mg/Fe]} & \colhead{[Ca/Fe]}}
\startdata
-1.60 &  0.60 &  0.40 &  0.32 \\
-1.40 &  0.50 &  0.40 &  0.30 \\
-1.20 &  0.50 &  0.40 &  0.28 \\
-1.00 &  0.40 &  0.40 &  0.26 \\
-0.80 &  0.30 &  0.34 &  0.26 \\
-0.60 &  0.20 &  0.22 &  0.17 \\
-0.40 &  0.20 &  0.14 &  0.12 \\
-0.20 &  0.10 &  0.11 &  0.06  \\
 0.00 &  0.00 &  0.05 &  0.00 \\
 0.20 &  0.00 &  0.04 &  0.00 
\enddata
\vspace{0.1cm} 
\tablecomments{We assume that Si and Ti trace Ca and that all other
  elements have [X/Fe]$=0.0$ for all metallicities.  See Section
  \ref{s:libabund} for details.}
\label{t:lib}
\end{deluxetable}

\subsection{Empirical SSPs}
\label{s:emp}

\subsubsection{New models and their behavior}

The construction of empirical SSPs follows standard isochrone
synthesis techniques \citep[for a review, see][]{Conroy13b}.  We use
the new MIST stellar evolution database for isochrones \citep{Choi16,
  Dotter16}.  These models cover all relevant evolutionary phases,
from the pre-main sequence through the post-AGB for lower mass stars,
and through the end of carbon burning for massive stars.  The
metallicity coverage ($-4.0\leq$[Fe/H]$\leq0.5$) is more than
sufficient for our purposes.  The isochrones compare favorably with a
variety of observational constraints \citep{Choi16}.  Improvements and
extensions to the MIST models are currently underway (including
$\alpha$-enhanced compositions) and will be presented in Dotter et
al. in preparation.

Each isochrone point is assigned a stellar spectrum by utilizing a
polynomial spectral interpolator presented in \citet{Villaume17a}.  The
interpolator was constructed from the MILES and Extended IRTF (E-IRTF)
libraries, with stellar parameters measured from the MILES spectra
presented in \citet{Prugniel11} and \citet{Sharma16}.  The E-IRTF
library consists of 284 stars carefully selected to span the parameter
space of $\teff$, $\logg$, and metallicity relevant for constructing
stellar population models of old stars.  The 284 E-IRTF stars along
with their optical spectra from MILES form the basis for the
interpolator.  The interpolator is supplemented with empirical spectra
from the M dwarf library of \citet{Mann15} as the MILES library
contains few very cool dwarfs.  In addition, theoretical spectral
models (built in the same way as the models presented in Section
\ref{s:rfn}) were employed to provide denser coverage of regions of
parameter space that are sparsely sampled by the empirical spectra.
The interpolator defines a smooth polynomial function at each
wavelength in the space of $\teff$, $\logg$, and metallicity for five
separate classes of stars.  The end result is a simple function that
interpolates empirical spectra to an arbitrary point in stellar
parameter space.

As with any interpolator, it is only as good as the input data, and so
one must be careful with regions of parameter space that are currently
lacking adequate coverage, especially at the metal-poor end \citep[see
Fig. 16 in][]{Villaume17a}.  In particular, at the lowest metallicities
([Fe/H]$<-1$) only the oldest isochrones are populated with empirical
spectra; younger models are entirely supported by the theoretical
spectral models and should therefore be used with greater caution.

The final ingredient for constructing empirical SSPs is the IMF.
Unless otherwise noted, all of the models presented in this paper were
constructed with a \citet{Kroupa01} IMF with lower and upper mass
cutoffs of $0.08\Msun$ and $100\Msun$, respectively.  Elsewhere
\citep[e.g.,][]{Conroy17a, Newman17, vanDokkum17a, Villaume17b} we have
presented models with more flexible IMFs, including Kroupa-like IMFs
with variable power-law indices and/or variable low-mass cutoffs, and
non-parametric IMFs, in which the weight within each $0.1\Msun$ bin is
a free parameter.

Owing to the different instrumental resolutions of the MILES and IRTF
libraries (FWHM$=2.5$\AA\, for MILES, corresponding to $R\approx2100$
at $0.5\mu m$, and $R\approx 2000$ for IRTF) we decided to smooth the
empirical SSPs to a common dispersion of $\sigma=100\kms$.  All other
model ingredients are also smoothed to this same resolution.

\begin{figure}[!t]
\center
\includegraphics[width=0.45\textwidth]{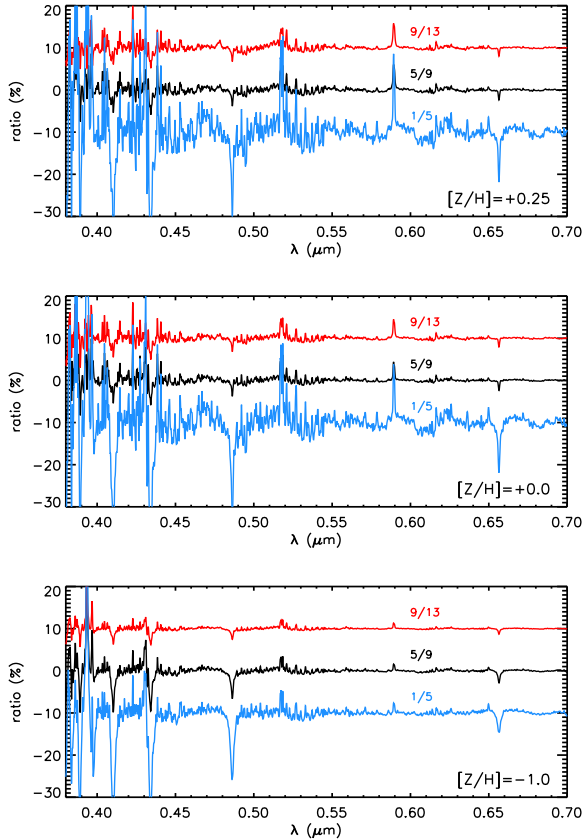}
\vspace{0.1cm}
\caption{Effect of age on empirical SSPs in the optical.  Each line is
  a ratio of two stellar ages (labeled in each panel as $t_1/t_2$).
  The ratio has been continuum-normalized to highlight the
  age-sensitive spectral features.  Each panel shows ratios at a
  different metallicity.}
\label{fig:sspage}
\end{figure}

\begin{figure}[t!]
\center
\includegraphics[width=0.45\textwidth]{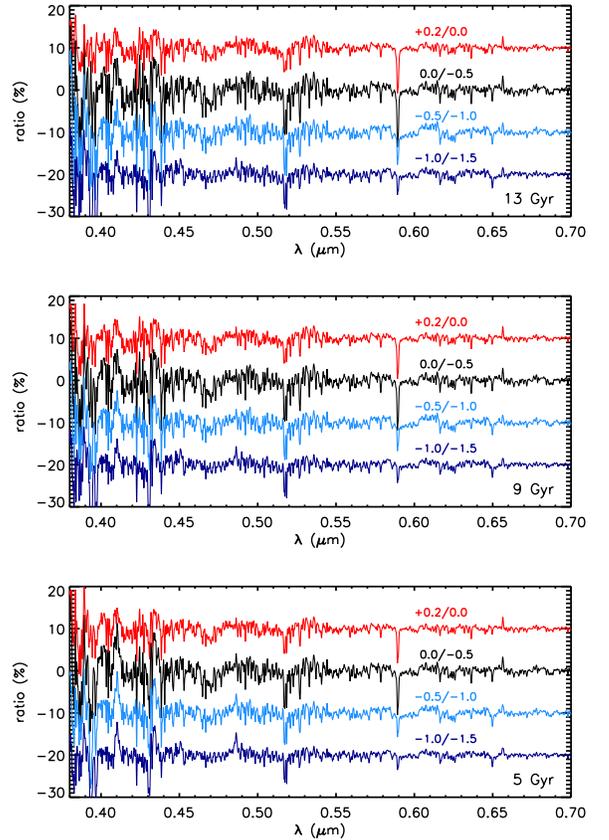}
\vspace{0.1cm}
\caption{Effect of metallicity on empirical SSPs in the optical.  Each
  line is a ratio of two metallicities (labeled in each panel as
  [Z/H]$_1$/[Z/H]$_2$).  The ratio has been continuum-normalized to
  highlight the metallicity-sensitive spectral features. Each panel
  shows ratios at a different age.  Note that the S07 models do not
  include predictions for NaD nor TiO$_1$.}
\label{fig:sspzmet}
\end{figure}

\begin{figure*}[t!]
\center
\includegraphics[width=0.9\textwidth]{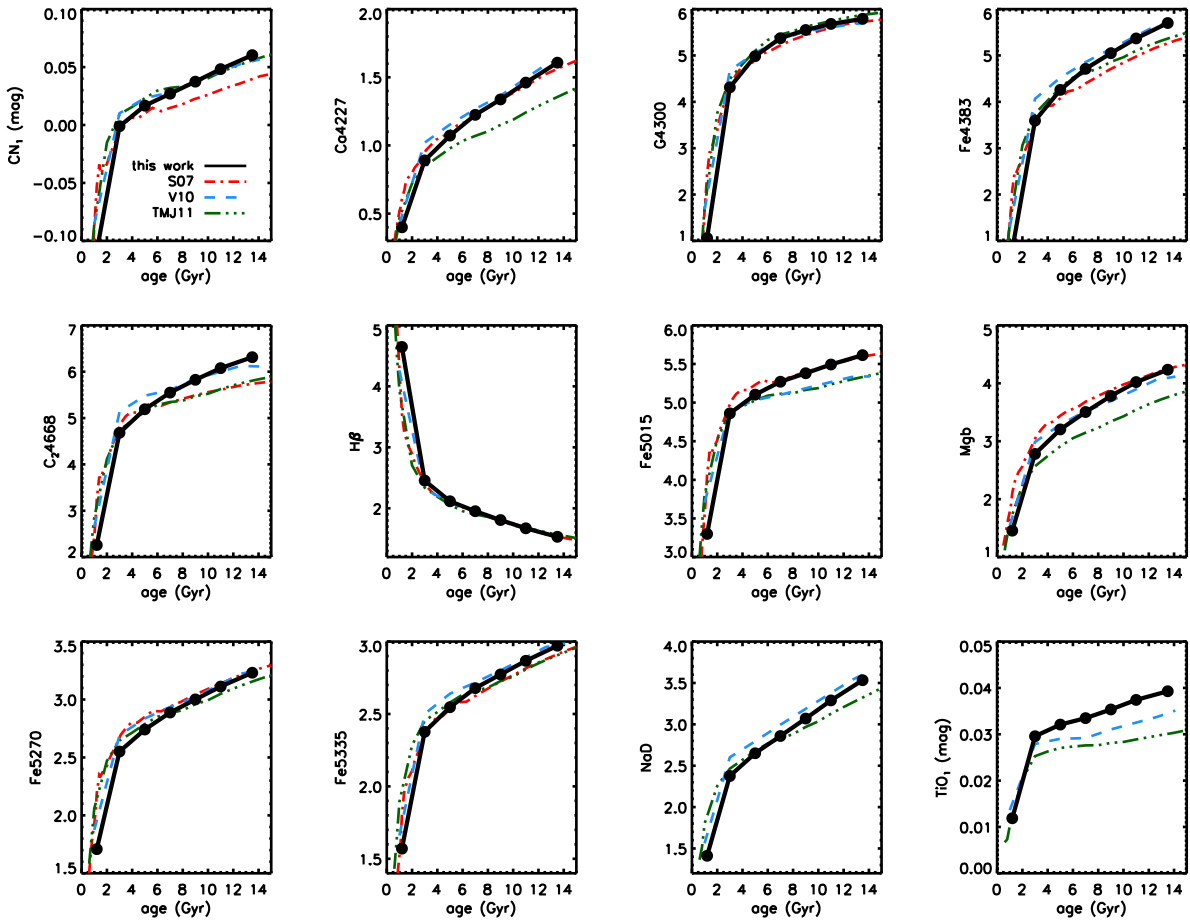}
\vspace{0.1cm}
\caption{Lick indices vs. age at [Z/H]=0.0.  Models presented in this
  work are compared to the models of \citet[][V10]{Vazdekis10},
  \citet[][S07]{Schiavon07}, and \citet[][TMJ11]{Thomas11}.  Indices
  are measured as equivalent widths in units of \AA\, unless otherwise
  noted.  Note that the S07 models do not include predictions for NaD
  nor TiO$_1$.}
\label{fig:lickage}
\end{figure*}

\begin{figure*}[t!]
\center
\includegraphics[width=0.9\textwidth]{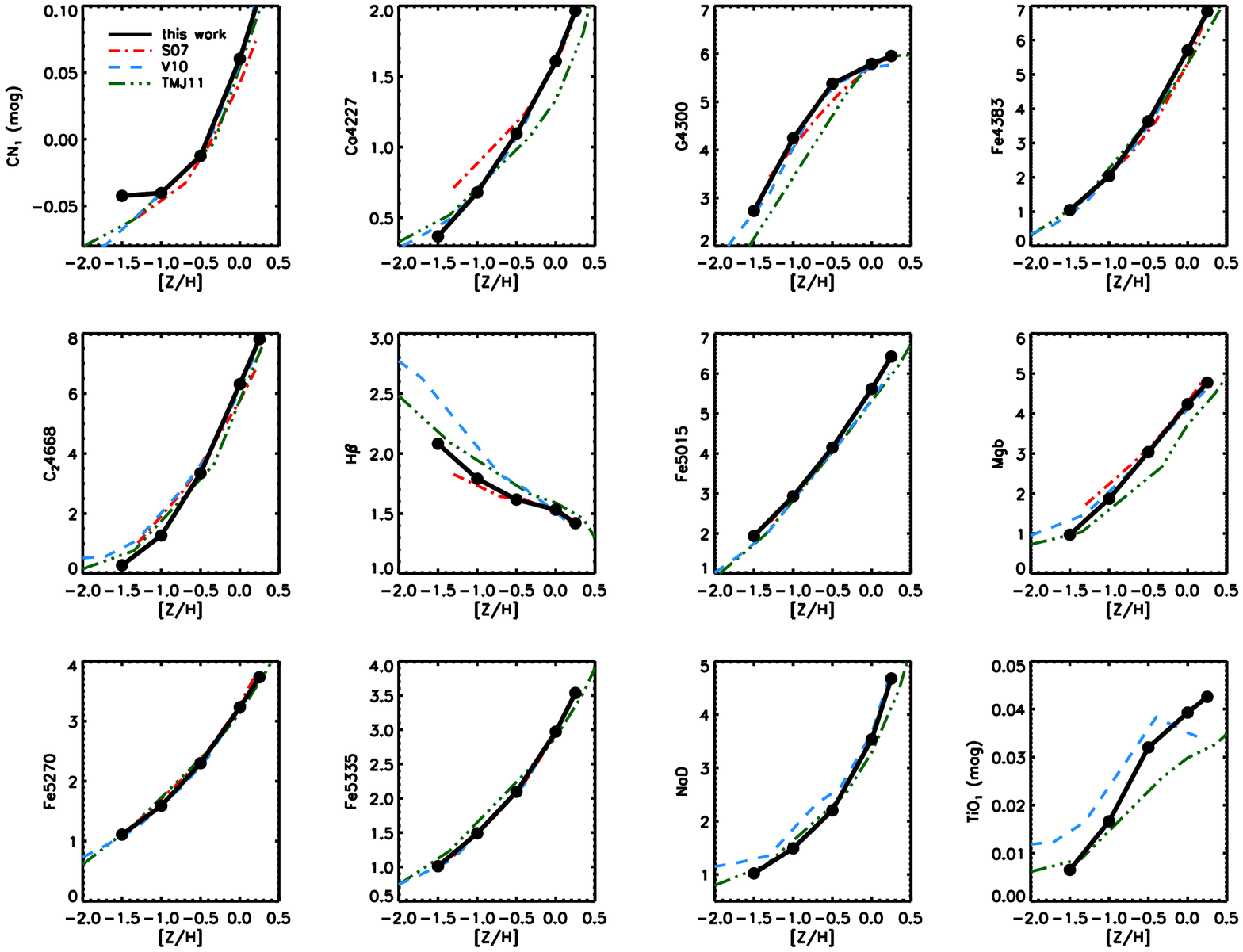}
\vspace{0.1cm}
\caption{Lick indices vs. metallicity at 13 Gyr.  Models presented in
  this work are compared to the models of \citet[][V10]{Vazdekis10},
  \citet[][S07]{Schiavon07}, and \citet[][TMJ11]{Thomas11}.  The S07
  and V10 models are shown at 14.1 Gyr as models at 13 Gyr are not
  available.  Indices are measured as equivalent widths in units of
  \AA\, unless otherwise noted.}
\label{fig:lickzmet}
\end{figure*}

In Figure \ref{fig:sspsum} we show the behavior of these new models as
a function of metallicity at 13 Gyr.  One clearly sees the two
principal effects of decreasing metallicity: decreasing line
strengths, due to the lower abundance of heavy elements, and a bluer
continuum, due both to hotter temperatures associated with metal-poor
isochrones and bluer SEDs at fixed temperature associated with less
line-blanketing opacity.  Note the gaps at $1.32-1.41\mu m$ and
$1.82-1.94\mu m$ due to poor atmospheric transmission.

In Figures \ref{fig:sspage} and \ref{fig:sspzmet} we highlight the
sensitivity of spectral features to changes in age and metallicity in
the $0.38-0.70\mu m$ window optical (the NIR wavelength range is shown
in Appendix C).  Figure \ref{fig:sspage} shows ratios of models of
different ages at fixed metallicities (one metallicity per panel).
The ratio has been divided by a polynomial to focus attention on the
behavior of the narrow spectral features.  As has been known for
several decades \citep[e.g.,][]{Worthey94}, the Balmer lines show very
strong sensitivity to age as they are temperature-sensitive.  Other
age-sensitive features include the \ionn{Mg}{i} triplet at $0.52\mu m$
and the NaD doublet at $0.59\mu m$.

Figure \ref{fig:sspzmet} shows ratios of models of different
metallicities at fixed ages (one age per panel).  We see strong
sensitivity to metallicity throughout the spectrum due to the numerous
absorption lines from many elements and molecules including Fe, Ti,
Mg, Na, CH, CN, and MgH.  Notice that the Balmer lines show much less
pronounced sensitivity to metallicity compared to age, a fact which
has long been exploited to break the so-called age-metallicity
degeneracy \citep[e.g.,][]{Worthey94}. The similarity of the ratio
spectra for different metallicity combinations is striking, in spite
of the fact that the overall absorption depths are of course much
weaker at lower metallicity.  We interpret this as a manifestation of
the fact that most transitions are on the linear part of the curve of
growth, so that a doubling of the species abundance results in a
doubling of the line equivalent width, irrespective of the absolute
line depth.

As noted in the Introduction, most previous work has focused on
modeling the equivalent widths (i.e., indices) of selected absorption
features.  In Figures \ref{fig:lickage} and \ref{fig:lickzmet} we
present 12 commonly-measured Lick indices \citep{Worthey94b} as a
function of age and metallicity.  We have measured the indices at the
Lick/IDS resolution \citep{Worthey97} in order to compare to other
models.  In these figures we compare our model predictions to
published models, including \citet[][S07]{Schiavon07},
\citet[][V10]{Vazdekis10}, and \citet[][TMJ11]{Thomas11}.  The S07
models use the empirical library from Jones et al. (1999) while the
models of V10 and TMJ11 utilize the MILES library.  S07 and TMJ11
present tabulated Lick indices at the Lick/IDS resolution, while we
have measured Lick indices (convolved to the Lick/IDS resolution for
fair comparison to the other models) from the V10 spectra directly.
All models in these figures were constructed with a \citet{Salpeter55}
IMF because the S07 and TMJ11 models were tabulated assuming this IMF.
V10 and S07 employ the \citet[][i.e., ``Padova'']{Girardi00}
isochrones, while TMJ11 employ the models from \citet[][i.e.,
``BaSTI'']{Cassisi97b}.  At least in principle, the difference between
the V10 and S07 models should only be due to stellar libraries (Jones
vs. MILES), while the differences between V10, TMJ11 and our own
models should reflect differences in the isochrones (Padova vs. BaSTI
vs. MIST).  In practice there are other issues to consider, including
how the empirical libraries are pruned for unreliable data, the source
of the stellar parameters, and how one interpolates within a sparely
and irregularly sampled parameter space.  In spite of all these
issues, the overall agreement between the four models is encouraging.

\begin{figure}
\center
\includegraphics[width=0.45\textwidth]{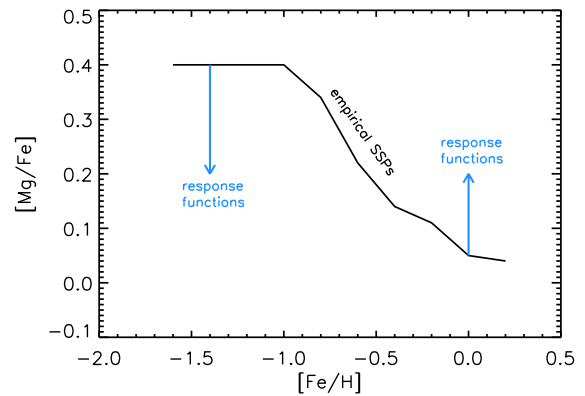}
\vspace{0.1cm}
\caption{Schematic demonstration of the interplay between the
  empirical SSPs and the response functions in determining the final
  element abundance.  The empirical SSPs inherit an abundance pattern
  from the underlying stellar library (black line, see Section
  \ref{s:libabund} for details).  In order to create models at
  arbitrary abundance pattern the response functions are used to
  correct the element abundance to an arbitrary value (blue arrows,
  see Section \ref{s:rfn} for details).  The example shown here is for
[Mg/Fe] where the black line represents the library abundance pattern
listed in Table \ref{t:lib}. }
\label{fig:cartoon}
\end{figure}

\begin{figure*}[t!]
\center
\includegraphics[width=0.9\textwidth]{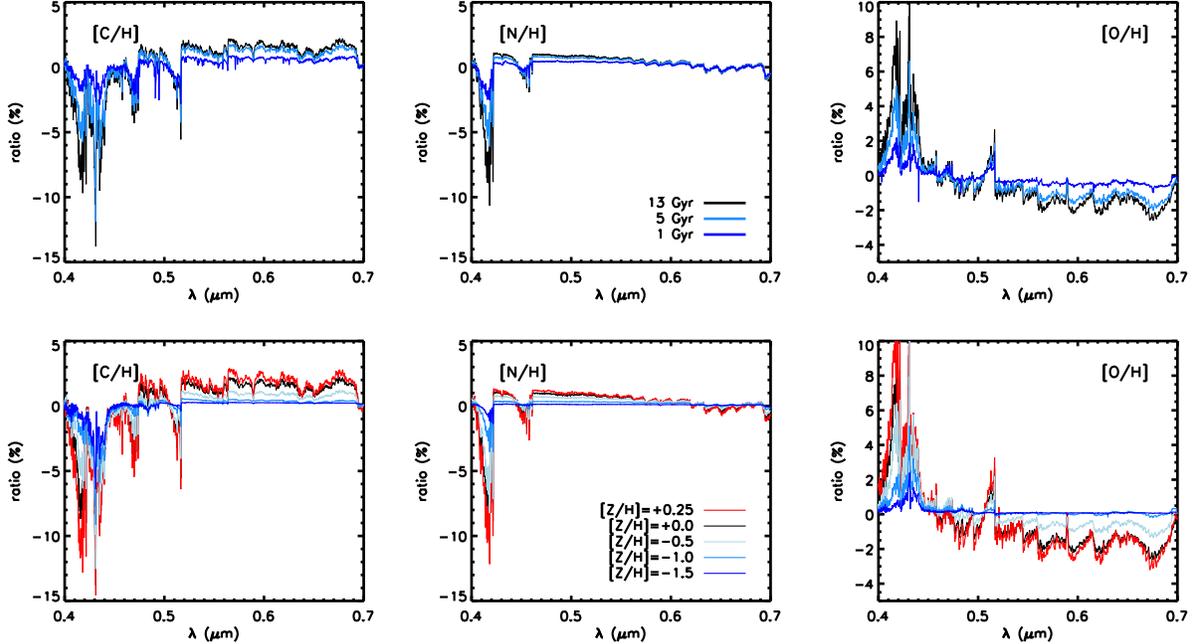}
\vspace{0.1cm}
\caption{Individual element response functions for CNO as a function
  of age and metallicity.  Response functions are the ratio between
  two theoretical SSPs: one that has an enhanced abundance of a single
  element, and another that has solar-scaled abundance ratios.  All
  enhancements are $+0.3$ dex except for carbon which is enhanced by
  $+0.15$ dex.}
\label{fig:cno}
\end{figure*}

\begin{figure*}[t!]
\center
\includegraphics[width=0.9\textwidth]{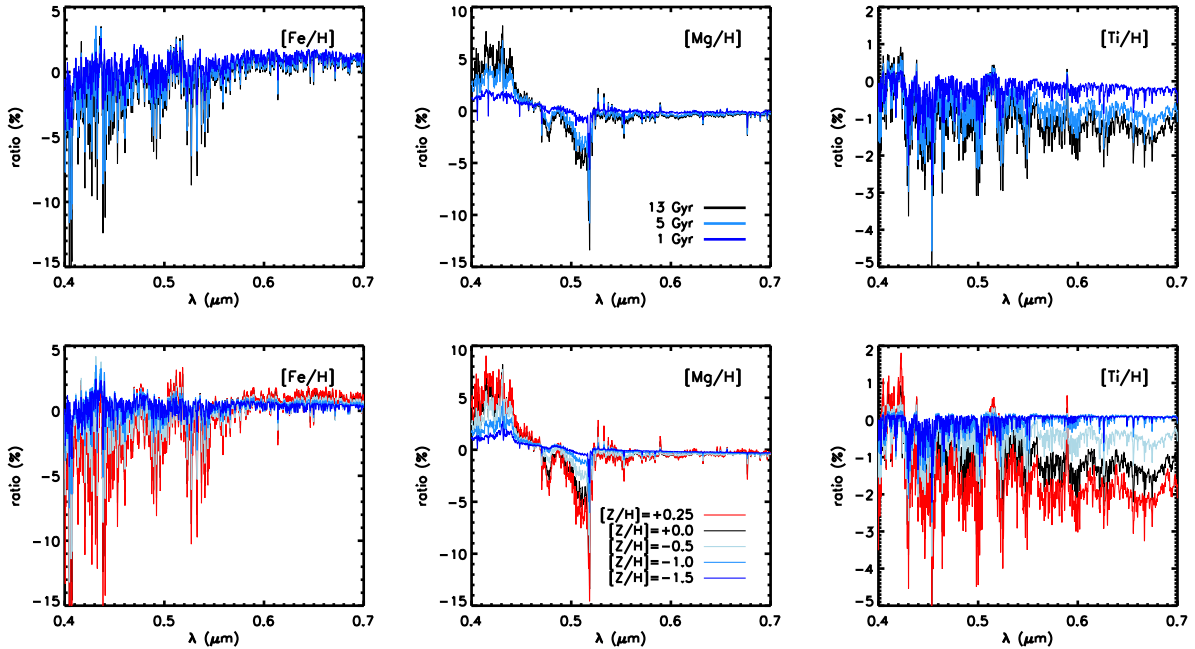}
\vspace{0.1cm}
\caption{Same as Figure \ref{fig:cno}, now for Fe, Mg, and Ti. }
\label{fig:femgti}
\end{figure*}

\subsubsection{Abundance pattern of the library stars}
\label{s:libabund}

In addition to overall metallicity, one must consider changes in
individual elements when constructing stellar population models.
Ideally, there would be uniform coverage in abundance for every
element of interest, and models could then be constructed from
empirical spectra for arbitrary abundance patterns.  In practice this
is not possible, both because the number of required stars would be
enormous, and because in practice Galactic chemical evolution limits
the range of element abundances at a given overall metallicity.  This
is why theoretical spectra are often used to enable abundance
variation in the models, as described in the following section.  For a
schematic illustration of the issue, see Figure \ref{fig:cartoon}.
However, first, the abundance patterns of the library stars must be
known because this sets the overall abundance scale at each
metallicity.

This turns out to be a challenging task, as the majority of MILES
stars do not have corresponding high resolution spectra from which
detailed element abundance determination has been performed.
\citet{Milone11} recently reported [Mg/Fe] abundances for 76\% of the
MILES stars, derived either from a literature compilation or their
own analysis of the MILES data.  \citet{Vazdekis15} recently used
these abundances to present the first empirical SSPs with
[$\alpha$/Fe] variation.

A standard approach is to tabulate the average abundance for each
element as a function of metallicity, and to then fold this
information into the derivation of abundances from the models so that
the final abundance patterns are on an absolute scale
\citep[e.g.,][]{Thomas03, Schiavon07, Thomas11}.   In practice this
means that the abundance of element X is derived via:
\begin{equation}
\label{e1}
\rm{[X/Fe]} =\rm{[X/Fe]}_{\rm lib}(\rm{[Z/H]}) + \rm{[X/Fe]}_{\rm resp. func.}, 
\end{equation}
\noindent
where the first term represents the (metallicity-dependent) abundance
pattern of the empirical library, and the second term represents the
correction determined from the theoretical response functions.  The
first term is described in this section while the second term is
described in Section \ref{s:rfn}.

In Table \ref{t:lib} we tabulate the adopted abundance patterns as a
function of metallicity for O, Mg, and Ca.  For [O/Fe] we adopt the
relation from \citet{Schiavon07}, except that we force [O/Fe]$=0.0$
for [Fe/H]$\ge0.0$.  The adopted relation agrees with the stellar
abundances derived from high resolution spectra in \citet{Bensby14}.
The relation for [Mg/Fe] is derived from the parameters for MILES
stars reported in \citet{Milone11}, while the [Ca/Fe] relation is
derived from the results in \citet{Bensby14}.  Recent results from
high resolution spectral analysis suggest that the heavier $\alpha$
elements trace each other \citep{Bensby14} and so we assume that Si
and Ti have the same library abundance pattern as Ca.

The Fe-peak elements largely trace Fe \citep[e.g.,][]{Bensby14,
  Holtzman15}, and so we assume [X/Fe]$=0.0$ for V, Cr, Mn, Co, Ni,
and Cu.  [Na/Fe] also appears to be close to 0.0 over a wide
metallicity range \citep{Kobayashi06, Bensby14, Holtzman15} and so we
assume no library enhancement for Na.  For all other elements we
assume no library enhancement.  This is a minimal assumption, though
it is important to note that it may result in non-trivial systematic
uncertainties at lower metallicities.  For example, for C and N there
is no strong evidence for variation with metallicity \citep[see
literature compilation in][]{Kobayashi06}, but the data are
particularly sparse for these elements.

\subsection{Theoretical Response Functions}
\label{s:rfn}

\subsubsection{New models and their behavior}

As noted in the previous section, given the limited dynamic range in
abundance pattern for the empirical spectra, we must turn to
theoretical models to make predictions for arbitrary abundances.  Our
basic approach is similar to previous work \citep[e.g.,][]{Tripicco95,
  Korn05, Serven05, LeeHC09}, and is presented in greater detail in
\citet{Conroy12a}.  The basic idea is to generate response functions,
which quantify the fractional change in the spectrum due to a change
in a single element abundance.  One then obtains arbitrary abundance
patterns by multiplying response functions together for all the
elements of interest (see Figure \ref{fig:cartoon} for a schematic
illustration).

We use the Kurucz suite of routines for computing model atmospheres
and spectra \citep{Kurucz70, Kurucz81, Kurucz93}.  A new model
atmosphere and spectrum is computed for every abundance change.  The
atmospheres are plane-parallel and hydrostatic.  The standard
mixing-length theory is adopted for the treatment of convection, and
local thermodynamic equilibrium (LTE) is assumed.  Spectra were
computed at $R=300,000$ from $0.3\mu m$ to $2.4\mu m$ with a fixed
microturbulence of $2\kms$ \citep[see][for a discussion of the effect
of microturbulence on the response functions]{Tripicco95, Conroy12a}.
We use the latest atomic line list from R. Kurucz (priv. comm.) and a
comprehensive set of molecules including H$_2$O, TiO, MgH, CH, CN,
MgO, AlO, NaH, VO, FeH, H$_2$, NH, C$_2$, CO, OH, SiH, SiO, CrH, and
CaH.  The spectra were then smoothed to $\sigma=100\kms$ and
downsampled to the empirical library wavelength grid.  We adopt the
latest solar abundances from \citet{Asplund09}, which is also the
abundance scale used for computing the MIST isochrones.

The atomic and molecular line lists are under a near-constant state of
improvement.  R. Kurucz has recently completed a comprehensive
revision of his atomic line list, including new calculations and
incorporation of new laboratory measurements.  We are frequently
updating our models to use the latest line lists available from
Kurucz.  In addition, Kurucz produces line lists for relevant
molecules based on the latest laboratory data and ab initio electronic
structure calculations.  Recent additions include FeH
\citep{Dulick03}, CaH \citep{Yadin12, Shayesteh13}, VO
\citep{McKemmish16}, and an updated SiH line list.  See
\texttt{http://kurucz.harvard.edu/molecules.html} for the full list.
The atomic and molecular line lists harbor non-trivial uncertainties
in the wavelengths, oscillator strengths, and damping constants for
many lines.  It is challenging to quantify the impact of these
uncertainties on the construction of stellar population models, which
is why the globular cluster tests presented in Section \ref{s:res} are
critical in that they provide ``end-to-end'' tests of the models.

Spectra were computed at $20-40$ points along each isochrone
for ages of 1, 3, 5, 9, and 13.5 Gyr, and for
[Fe/H]$=-1.5, -1.0, -0.5, +0.0, +0.2$.  SSPs were constructed for the
reference abundance scale and then for each enhanced (or depressed)
element, and the final response functions were computed by taking
ratios between the SSPs with altered and solar-scaled abundances.  For
each element we computed altered abundance models with an enhancement
of [X/H]$=+0.3$, except for C where we constructed models with
[C/H]$=+0.15$ in order to avoid the formation of carbon stars.  For N
and Na we created additional, higher enhancement models, as
observations routinely require enhancements in excess of +0.3 dex.  In
addition, for the most important elements we also created depressed
models, with [X/H]$=-0.3$.

It is not immediately obvious that creating arbitrary abundance
pattern models through linear combinations of individual response
functions should lead to accurate predictions.  In practice, the
abundance shifts are relatively modest, so that the effect on the
spectrum can be thought of as a Taylor expansion to first or second
order.  In that limit the higher-order effects on the spectrum (and
atmosphere) due to two or more elements varying simultaneously are minor
\citep[see][for detailed discussion of this point]{Ting16b}.  We have
explicitly tested this assumption in a variety of contexts and find
that the standard assumption is remarkably robust.

Example response functions are shown in Figures \ref{fig:cno} and
\ref{fig:femgti} for the elements C, N, O, Fe, Mg, and Ti in the
optical wavelength range (the NIR wavelength range is shown in 
Appendix C).  In each figure the upper set of panels show the behavior
of the response function vs. age and the lower panels show the
behavior vs. metallicity.  We focused on specific wavelengths of
interest though we emphasize that there is useful information over the
full wavelength range considered for many elements.  Many of the
strongest features in these figures are due to molecules, including
CH, CN, TiO, and MgH.  The behavior of the oxygen response function in
the blue is mainly due to the indirect effect of oxygen on CNO
molecular equilibrium and atmospheric effects. At redder wavelengths
the response function is dominated by TiO.

\begin{figure}[!t]
\center
\includegraphics[width=0.49\textwidth]{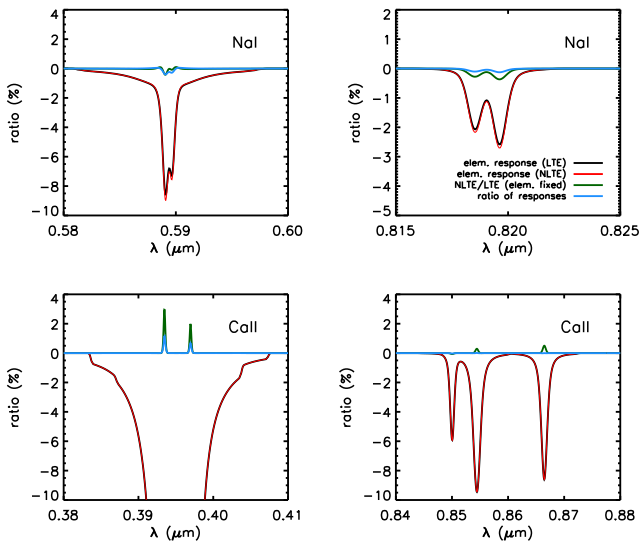}
\vspace{0.1cm}
\caption{Effect of LTE vs. NLTE on the theoretical response functions
  for a 13 Gyr solar metallicity SSP. In each panel we compare the
  element response functions for LTE (black lines) and NLTE (red
  dashed lines).  We also show the ratio of these response functions
  (blue lines) and the ratio of NLTE to LTE models with fixed
  abundances (green lines).  The green lines show the effect of LTE
  vs. NLTE on the direct model predictions, while the blue lines show
  the effect of LTE vs. NLTE on the element response functions.  In
  effect, the blue lines represent the (very small) systematic errors
  we are introducing by assuming LTE to compute the element response
  functions.}
\label{fig:nlte}
\end{figure}

\subsubsection{NLTE effects}

It is well-known that departures from LTE can substantially alter the
depths and shapes of absorption lines, especially at low surface
gravity and/or low metallicity \citep[e.g.,][]{Asplund05}.  The effect
of NLTE can be important even at low spectral resolution at solar
metallicity \citep{Short15}.  However, in our models, like many others
\citep{Trager00, Thomas03, Schiavon07, Thomas11}, the theoretical
spectra are only used differentially via the response functions, and
so it unclear to what extent NLTE may play a role.

In order to explore this issue, we have computed LTE and NLTE spectral
synthesis models for the two strongest Na and Ca lines in the
$0.37-1.0\mu m$ range.  These calculations rely on the model atom for
Na presented in \citet{Lind11} and for Ca from \citet{Melendez17}.
The LTE and NLTE models were computed with the MULTI code
\citep{Carlsson86} along with MARCS stellar atmospheres
\citep{Gustafsson08}.  The reader is referred to \citet{Lind11} for
details regarding the modeling procedure.  LTE and NLTE models were
computed for a 3500K dwarf, a 3500K giant, and a 5500K turnoff star.
Model SSPs were constructed from these three stars.  All models have
been velocity broadened to $100\kms$ to match the resolution of the
empirical SSPs.

The results are shown in Figure \ref{fig:nlte}.  In each panel we
explore the effect of NLTE on the element response functions at solar
metallicity and 13 Gyr.  We compare the response functions for a 0.3
dex element enhancement computed with LTE and NLTE assumptions (black
and red lines).  The ratio of these two lines is shown as the blue
line.  This represents the relative error introduced from the
assumption of LTE in the response functions.  Because the response
functions generally result in several to tens of percent change in the
spectrum, the {\it absolute} error associated with the LTE assumption
is even smaller.  The key result is that this relative error is very
small, $\ll 1$\% in all cases except for the cores of the
\ionn{Ca}{ii} H\&K lines where the effect reaches $\approx1$\% (note
that we generally avoid including the H\&K lines when fitting high
quality absorption line spectra).  This result can be understood as a
consequence of the fact that the curves of growth for LTE and NLTE
lines are similar in shape.  Response functions are sensitive to the
{\it slope} of the curve of growth since they are computed from ratios
of models, and so similarly-shaped curves of growth should produce
similar response functions.

We also plot the ratio of LTE to NLTE spectra for a fixed abundance
(green lines).  This would be the error incurred if we were to use the
synthetic models in an absolute sense rather than differentially.
Comparison between the blue and green lines highlights the advantage
of working with synthetic spectra in a differential sense - the error
incurred from the assumption of LTE is smaller.

The conclusion to draw from this section is that our assumption of LTE
in the construction of theoretical response functions is adequate at
the sub-percent level.  We have only tested the effect of NLTE on a
handful of transitions, but given that these lines span a range of
strengths, including some with saturated cores where NLTE effects tend
to be strongest, we can anticipate that NLTE effects will generally be
weak in response functions especially when convolved to a resolution
of $\sigma\gtrsim100\kms$.


\begin{deluxetable}{llc}
\tablecaption{Summary of Fitting Parameters}
\tablehead{ \colhead{Parameter} & \colhead{Description} & \colhead{Range} }
\startdata
{\it simple mode:}\\
$v_z$ & Recession velocity & N/A \\
$\sigma$ & Velocity dispersion ($\kms$)  &  ($10,10^3$) \\
$t_o$  & Stellar population age (Gyr) & ($1,14$)$^{1}$ \\
Z/H  & Stellar population metallicity & ($-2.0,0.3$) \\
C/H   & Carbon abundance & ($-0.3,0.5$) \\
N/H   & Nitrogen abundance & ($-0.3,1.0$) \\
O/H   & Oxygen abundance & ($-0.3,0.5$) \\
Na/H   & Sodium abundance & ($-0.3,1.0$) \\
Mg/H   & Magnesium abundance & ($-0.3,0.5$) \\
Si/H   & Silicon abundance & ($-0.3,0.5$) \\
Ca/H   & Calcium abundance & ($-0.3,0.5$) \\
Ti/H   & Titanium abundance & ($-0.3,0.5$) \\
Fe/H   & Iron abundance & ($-0.3,0.5$) \\
\\
{\it full mode:}\\
$t_y$  & Age of young component (Gyr) & (1,3) \\
log($f_y$) & Mass fraction of young component & ($-6.0,-0.1$) \\
K/H   & Potassium abundance & ($-0.3,0.5$) \\
V/H   & Vanadium abundance & ($-0.3,0.5$) \\
Cr/H   & Chromium abundance & ($-0.3,0.5$) \\
Mn/H   & Manganese abundance & ($-0.3,0.5$) \\
Co/H   & Cobalt abundance & ($-0.3,0.5$) \\
Ni/H   & Nickel abundance & ($-0.3,0.5$) \\
Cu/H   & Copper abundance & ($-0.3,0.5$) \\
Sr/H   & Strontium abundance & ($-0.3,0.5$) \\
Ba/H   & Barium abundance & ($-0.3,0.5$) \\
Eu/H   & Europium abundance & ($-0.3,0.5$) \\
$\alpha_1$ & IMF slope from $0.08-0.5\Msun$ & (0.5,3.5) \\
$\alpha_2$ & IMF slope from $0.5-1.0\Msun$ & (0.5,3.5) \\
$m_c$ & IMF lower-mass cutoff ($\Msun$) & (0.08,0.4) \\
log($f_{\rm hot}$) & Hot star component & ($-6.0,-1.0$) \\
$T_{\rm hot}$ & $\teff$ of hot star component (kK) & (8,30) \\
$\sigma_e$ & Velocity dispersion of emission lines & ($10,10^3$) \\
$v_e$ & Relative velocity of emission lines & ($-10^3,10^3$) \\
log($e_H$) &  Amplitude of Balmer emission lines  & ($-6.0,1.0$) \\
log($e_{\rm [OIII]}$) &  Amplitude of [\ionn{O}{iii}] emission lines  & (-$6.0,1.0$) \\
log($e_{\rm [NI]}$) &  Amplitude of [\ionn{N}{i}] emission line  & ($-6.0,1.0$) \\
log($e_{\rm [NII]}$) &  Amplitude of [\ionn{N}{ii}] emission lines  & ($-6.0,1.0$) \\
log($e_{\rm [SiII]}$) &  Amplitude of [\ionn{S}{ii}] emission lines  & ($-6.0,1.0$) \\
$h_3$   & Gauss-Hermite parameter & ($-0.4,0.4$) \\
$h_4$   & Gauss-Hermite parameter  & ($-0.4,0.4$) \\
$j$   & Jitter term (error inflation) & ($0.1,10$) \\
log($f_{\rm sky}$) & Sky emission lines (error inflation) & ($-6.0,2.0$)\\
log($T_{\rm trans}$) & Sky transmission & ($-6.0,1.0$) 
\enddata
\vspace{0.1cm} 
\tablecomments{See Section \ref{s:fit} for explanation of the
  parameters.  Note that the allowed range for the element abundances
  refers to response function term in Equation 1.}
\tablenotetext{1}{In the full mode, the lower limit on the parameter $t_o$
is 3 Gyr in order to avoid degeneracy with the parameter $t_y$.}
\label{t:alf}
\end{deluxetable}

\section{Fitting Technique}
\label{s:fit}

In this section we describe our approach to fitting models to data.
This approach has been described in detail in \citet{Conroy12b} with
updates detailed in \citet{Conroy14a} and \citet{Choi14}.  As a number
of features have been added and updated in the past several years we
provide a summary of the key points.  Our code for implementing the
technique described below is called \alf, for ``absorption line
fitting''.  The models are available for download at
\texttt{https://scholar.harvard.edu/cconroy/sps-models}.

\subsection{Full Spectrum Fitting with \alf}
\label{s:fitting}

\begin{figure*}[!t]
\center
\includegraphics[width=0.9\textwidth]{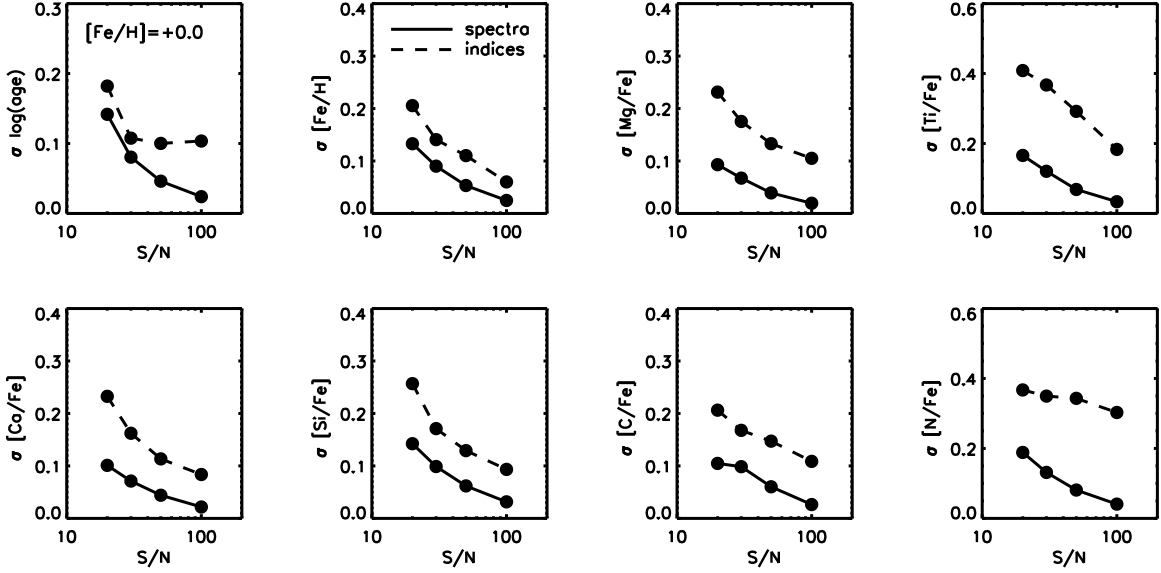}
\vspace{0.1cm}
\caption{Uncertainties on derived parameters as a function of S/N for
  mock data.  Mock data were constructed with an age of 10 Gyr,
  [Fe/H]$=+0.0$, and solar-scaled abundance patterns.  Solid lines
  show results when fitting the full optical spectrum from
  $0.40-0.56 \mu m$ while dashed lines show the errors derived from
  measuring and fitting 14 indices over the same wavelength range from
  the same spectra.}
\label{fig:indzp}
\end{figure*}

\begin{figure*}[!t]
\center
\includegraphics[width=0.9\textwidth]{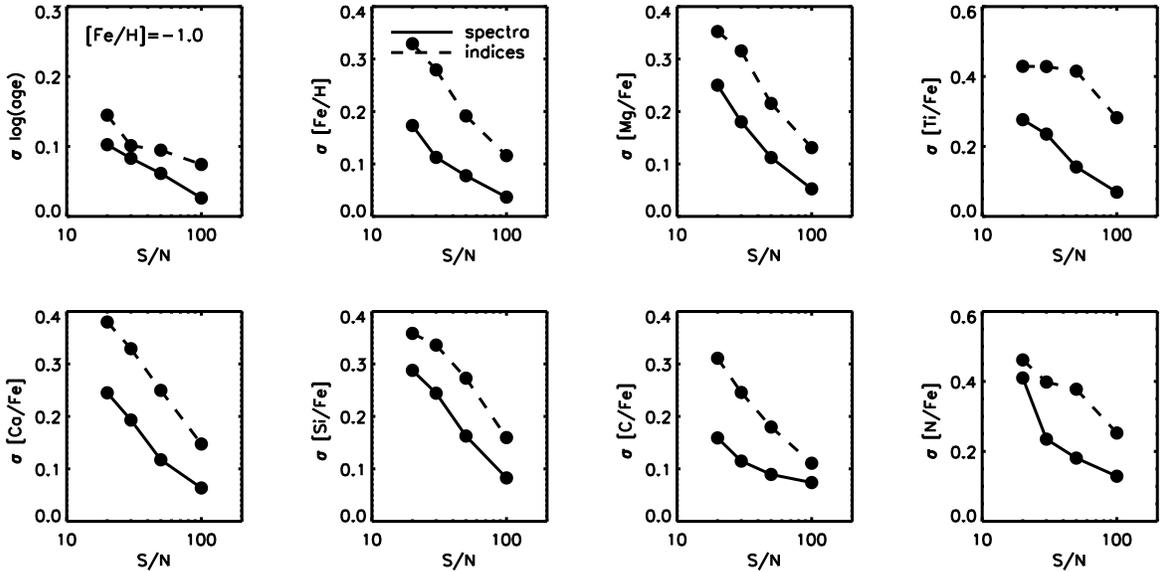}
\vspace{0.1cm}
\caption{Same as Figure \ref{fig:indzp}, except now for mock data with
[Fe/H]$=-1.0$.}
\label{fig:indzm}
\end{figure*}

\begin{figure}[!t]
\center
\includegraphics[width=0.5\textwidth]{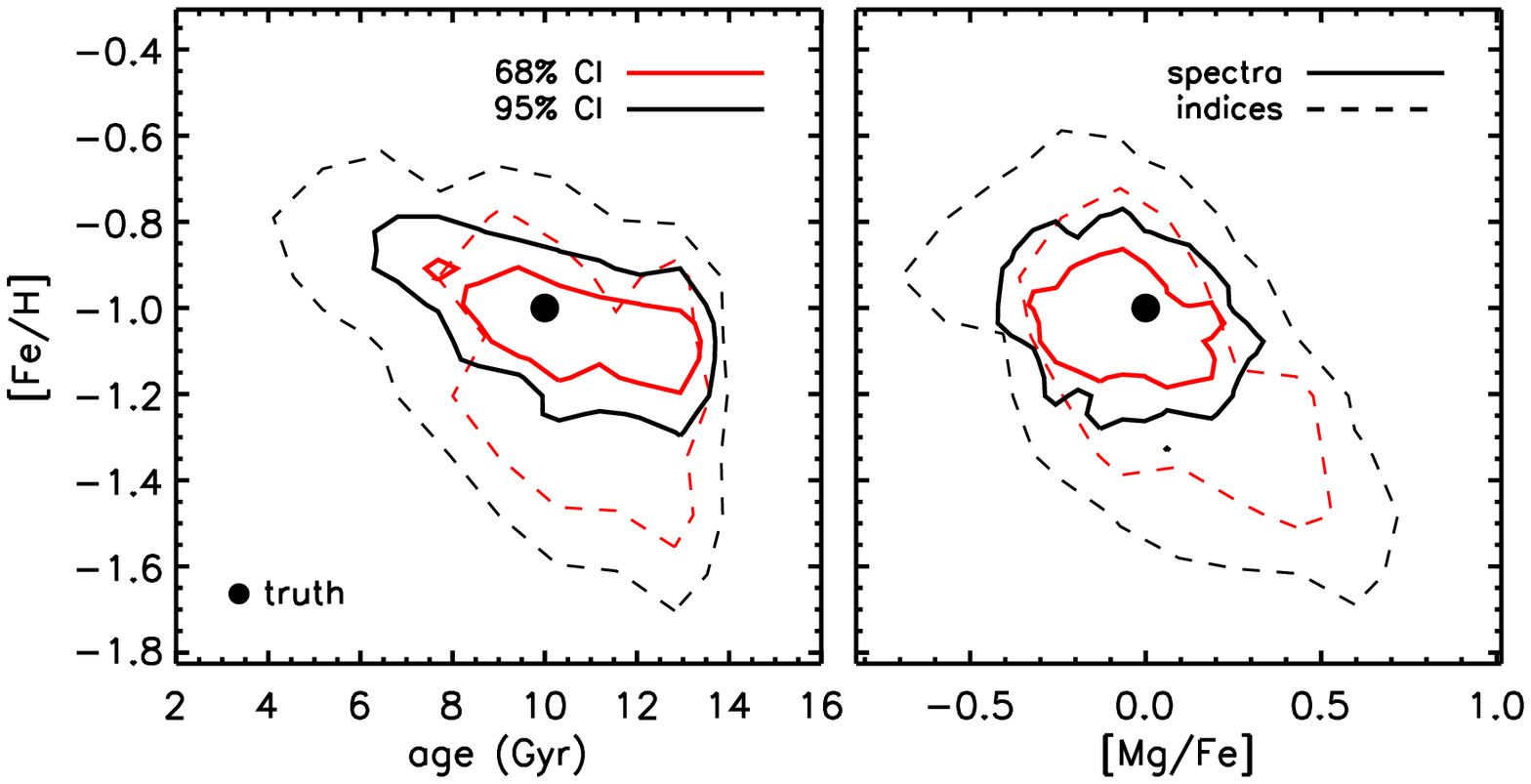}
\vspace{0.1cm}
\caption{Posterior probability distributions for age, [Fe/H], and
  [Mg/Fe] based on mock data with S/N$=30$.  Constraints from full
  spectrum fitting (solid lines) are compared to constraints from
  fitting spectral indices (dashed lines). }
\label{fig:corner}
\end{figure}

The fitting algorithm works in one of two modes, which we refer to as
``simple'' and ``full''.  In the simple mode 13 parameters are fit
including the recession velocity ($v_z$), velocity dispersion
($\sigma$), stellar age ($t_o$), Z/H, and response function-based
abundances of Fe, C, N, O, Na, Mg, Si, Ca, and Ti.  The IMF is fixed
to the \citet{Kroupa01} form.  This fitting mode is appropriate when
the data is either of low S/N or only a limited wavelength range
(e.g., blue optical) is available.  

The full mode includes these and an additional 29 parameters: a
younger age component with a variable age ($t_y$) and mass fraction
($f_y$); response function-based abundances of K, V, Cr, Mn, Ni, Cu,
Sr, Ba, and Eu; the logarithmic slopes of the IMF in the stellar mass
intervals $0.1-0.5\Msun$ ($\alpha_1$) and $0.5-1.0\Msun$ ($\alpha_2$),
and a free low-mass cutoff ($m_c$); an additional hot star component
with a variable $\teff$ ($T_{\rm hot}$) and flux contribution
($f_{\rm hot}$); a set of emission lines that can be strong in the
$0.40-0.70\mu m$ range including the Balmer lines (H$\delta$,
H$\gamma$, H$\beta$, H$\alpha$) with line ratios determined by
assuming Case B recombination \citep{Osterbrock89}, and
[\ionn{O}{iii}], [\ionn{N}{i}], [\ionn{N}{ii}], and [\ionn{S}{ii}].
These emission lines have a velocity dispersion independent of the
continuum ($\sigma_e$) and can be offset in velocity from the
continuum ($v_e$).  The stellar continuum is broadened by a
line-of-sight velocity distribution that includes two higher-order
moments parameterized by Gauss-Hermite polynomials with parameters
$h_3$ and $h_4$ \citep[see][for details]{vanderMarel93, Cappellari04}.
Finally, we include three parameters related to potential issues with
the data: a jitter term that allows us to scale up or down the quoted
errors by a multiplicative factor $j$; a term ($f_{\rm sky}$) added to
the quoted errors that scales with the location of bright sky lines
from the model of \citet{Noll12}; a term that applies atmospheric
absorption to the model ($T_{\rm trans}$) to account for the
possibility that the data were not (or was imperfectly) corrected for
telluric absorption.  The atmospheric absorption is based on the
models of \citet{Clough05}.

All 41 parameters included in \alf\, and their nominal allowed ranges
are listed in Table \ref{t:alf}.

This large parameter space is efficiently explored with Markov chain
Monte Carlo techniques.  We employ \texttt{emcee},
\citep{Foreman-Mackey13}, an affine-invariant ensemble sampler
\citep{Goodman10} that has been ported to Fortran.  This approach
allows us to derive the full posterior distributions and parameter
covariances.  We utilize 512 walkers and 20,000 burn-in steps before a
production run of 1,000 steps is used for the final posterior
distributions.  In our experience this setup results in chains that
are always well converged.  In this paper we quote best-fit values at
the 50\% credible interval (CI), i.e., the median of the distribution,
and errors as the 16\% and 84\% CIs.  In the simple mode of \alf\,
used herein, the likelihood is the usual: 
\begin{equation}
{\rm ln}\, \mathcal{L} \propto -0.5\, \sum_i (d_i-m_i)^2/\sigma_i^2
\end{equation}
where $d$ and $m$ are the data and model, respectively, $\sigma$ is
the uncertainty on the data, and the sum is over all wavelengths.

Tests with mock data are critical to ensure that the overall fitting
algorithm is robust, that the likelihood sampling algorithm has
converged to the true posteriors, and to understand covariances
between parameters in the posterior space.  We have presented mock
tests in \citet{Choi14} and \citet{Conroy17a} showing explicitly the
behavior of the derived parameters as a function of S/N.  Later in
this section we present additional mock tests focusing on the error
budget as a function of S/N and metallicity.

We approach model fitting by comparing to observations in spectral
space, rather than comparing in index space.  We match the continuum
shape of the data by multiplying the model by polynomials.  We do this
for several reasons.  First, relative flux calibration is much more
challenging on $\sim1000$\AA\, scales compared to $\sim10$\AA\,
scales, and so one would have to include additional parameters to
account for imperfections in the data reduction.  Second, dust will
affect the continuum in a smooth manner, so more parameters would be
required to account for dust.  Our approach is to fit the ratio of the
data and model by a high order polynomial where the order $n$ is
determined by $n\equiv (\lambda_{\rm max}-\lambda_{\rm min})/100$\AA\,
within each wavelength interval \citep[see also][]{Kelson00}.
Specifically, the polyomial is of the form
$p(\lambda)=\sum_{i=0}^n\,c_i(\lambda-\mu)^{i}$ where $\mu$ is the
mean wavelength of the region being fit.  This form ensures that the
polynomial is sufficiently flexible to account for continuum mis-match
issues but not so flexible that it could over-fit broad absorption
features.  The polynomial coefficients $c_i$ are re-fit during each
iteration and are determined by a simple least squares minimization.
We have experimented with this approach in a number of ways, including
changing the order of the polynomial and the method by which the
polynomial is applied to the model and find that the results are
generally insensitive to these choices.  In Appendix B we present new
tests of our approach to continuum filtering the data.

\begin{figure*}[!t]
\center
\includegraphics[width=0.9\textwidth]{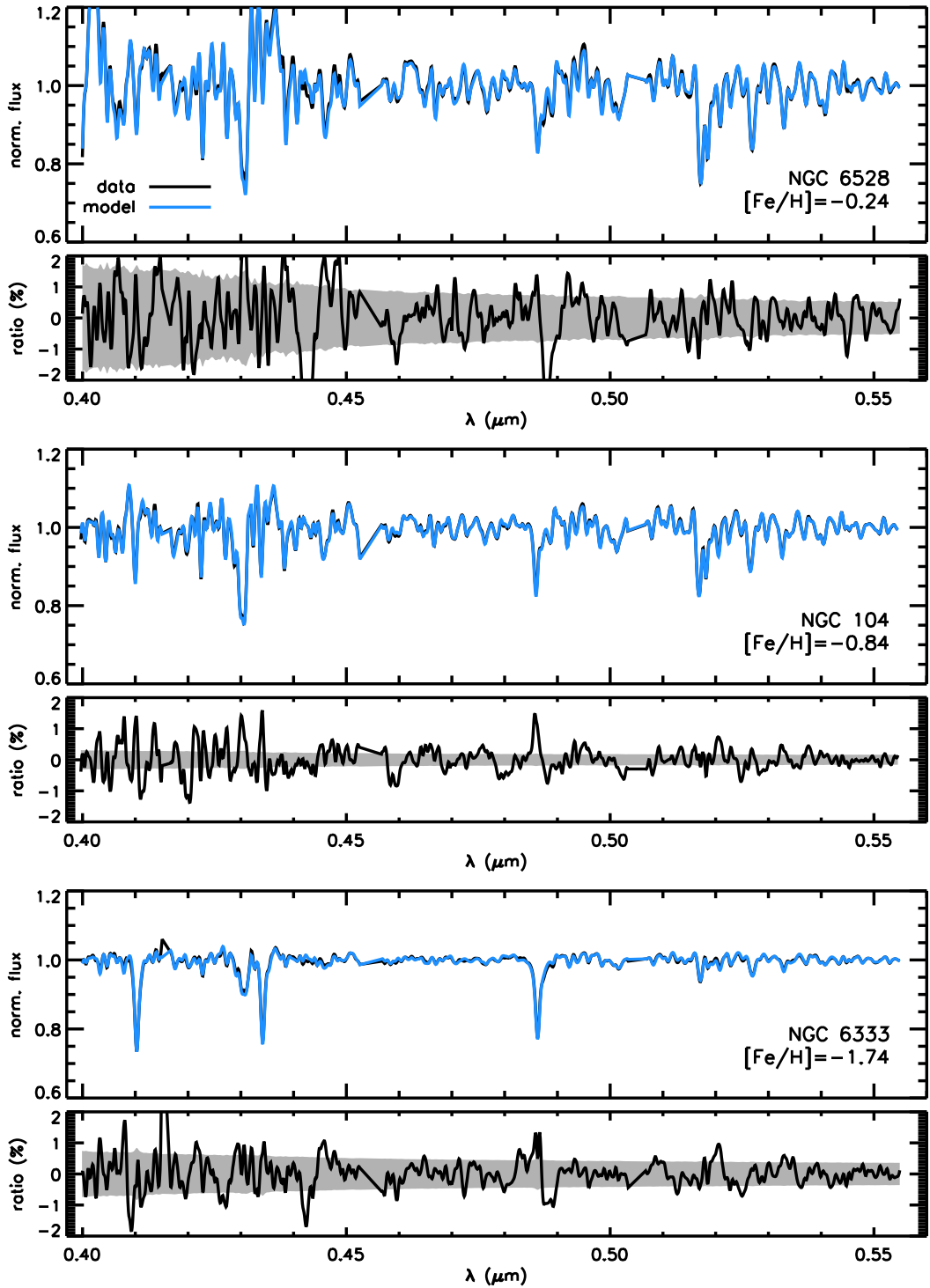}
\vspace{0.1cm}
\caption{Example fits to three GCs spanning a range of metallicities.
  The top panels show continuum normalized flux of the data and
  best-fit model.  Lower panels show the residuals between data and
  best-fit model (plotted as data/model$-1$) and the errors on the data
  (grey shaded bands).}
\label{fig:s05ex}
\end{figure*}

\begin{deluxetable*}{lrrrrrrrrrrrrrrrrrrrr}
\tablecaption{Derived Parameters of Galactic Globular Clusters}
\tablehead{ \colhead{NGC} & \colhead{Age} & \colhead{$\sigma_{\rm
      age}$} & \colhead{[Fe/H]} &
  \colhead{$\sigma_{\rm Fe}$} &
  \colhead{[C/Fe]} & \colhead{$\sigma_{\rm C}$} &\colhead{[N/Fe]} & \colhead{$\sigma_{\rm N}$}&
  \colhead{[Mg/Fe]} & \colhead{$\sigma_{\rm Mg}$}& \colhead{[Si/Fe]} &
  \colhead{$\sigma_{\rm Si}$} & \colhead{[Ca/Fe]} & \colhead{$\sigma_{\rm Ca}$} &
  \colhead{[Ti/Fe]} & \colhead{$\sigma_{\rm Ti}$}}
\startdata
      104 &  1.14 &  0.01 & -0.84 &  0.01 &  0.13 &  0.01 &  0.62 &  0.02 &  0.42 &  0.01 &  0.27 &  0.02 &  0.23 &  0.01 &  0.40 &  0.01 \\
    1851 &  0.89 &  0.01 & -1.19 &  0.02 & -0.11 &  0.04 &  0.31 &  0.06 &  0.29 &  0.03 &  0.19 &  0.05 &  0.19 &  0.04 &  0.31 &  0.03 \\
    1904 &  0.90 &  0.01 & -1.59 &  0.03 &  0.09 &  0.05 & -0.16 &  0.11 &  0.32 &  0.07 &  0.28 &  0.11 &  0.27 &  0.08 &  0.20 &  0.07 \\
    2298 &  0.86 &  0.03 & -1.69 &  0.08 &  0.20 &  0.09 & -0.03 &  0.23 &  0.52 &  0.17 &  0.24 &  0.16 &  0.36 &  0.18 &  0.32 &  0.16 \\
    2808 &  1.03 &  0.01 & -1.21 &  0.01 & -0.24 &  0.02 &  0.34 &  0.04 &  0.14 &  0.01 &  0.22 &  0.03 &  0.21 &  0.02 &  0.35 &  0.02 \\
    3201 &  0.84 &  0.02 & -1.33 &  0.03 &  0.07 &  0.04 & -0.11 &  0.12 &  0.24 &  0.04 &  0.11 &  0.05 &  0.16 &  0.07 &  0.30 &  0.06 \\
    5286 &  0.90 &  0.01 & -1.62 &  0.02 &  0.14 &  0.05 & -0.07 &  0.12 &  0.45 &  0.04 &  0.32 &  0.08 &  0.33 &  0.06 &  0.29 &  0.05 \\
    5904 &  0.83 &  0.01 & -1.32 &  0.01 &  0.06 &  0.06 &  0.10 &  0.07 &  0.44 &  0.02 &  0.26 &  0.05 &  0.28 &  0.03 &  0.22 &  0.03 \\
    5927 &  1.13 &  0.03 & -0.52 &  0.03 & -0.06 &  0.10 &  0.47 &  0.06 &  0.28 &  0.02 &  0.20 &  0.04 &  0.13 &  0.04 &  0.25 &  0.04 \\
    5946 &  0.89 &  0.03 & -1.54 &  0.06 &  0.08 &  0.08 &  0.10 &  0.26 &  0.34 &  0.09 &  0.26 &  0.16 &  0.48 &  0.15 &  0.22 &  0.12 \\
    5986 &  0.88 &  0.01 & -1.59 &  0.02 &  0.11 &  0.05 &  0.13 &  0.15 &  0.45 &  0.05 &  0.40 &  0.08 &  0.33 &  0.07 &  0.22 &  0.06 \\
    6121 &  0.85 &  0.01 & -1.23 &  0.02 &  0.00 &  0.06 &  0.31 &  0.08 &  0.53 &  0.03 &  0.23 &  0.07 &  0.30 &  0.04 &  0.28 &  0.05 \\
    6171 &  0.90 &  0.03 & -1.02 &  0.04 &  0.12 &  0.06 &  0.56 &  0.10 &  0.43 &  0.05 &  0.25 &  0.10 &  0.32 &  0.09 &  0.42 &  0.08 \\
    6218 &  0.85 &  0.02 & -1.43 &  0.03 &  0.16 &  0.06 &  0.01 &  0.16 &  0.61 &  0.05 &  0.35 &  0.09 &  0.40 &  0.07 &  0.22 &  0.07 \\
    6235 &  0.91 &  0.05 & -1.18 &  0.07 & -0.03 &  0.10 &  0.22 &  0.30 &  0.24 &  0.11 &  0.03 &  0.15 &  0.27 &  0.17 &  0.36 &  0.16 \\
    6254 &  1.00 &  0.01 & -1.56 &  0.03 &  0.09 &  0.05 & -0.16 &  0.12 &  0.36 &  0.06 &  0.31 &  0.10 &  0.39 &  0.06 &  0.30 &  0.06 \\
    6266 &  0.94 &  0.01 & -1.17 &  0.01 & -0.13 &  0.03 &  0.51 &  0.04 &  0.39 &  0.02 &  0.21 &  0.04 &  0.30 &  0.02 &  0.36 &  0.02 \\
    6284 &  0.84 &  0.01 & -1.25 &  0.03 &  0.07 &  0.05 &  0.61 &  0.09 &  0.51 &  0.04 &  0.32 &  0.07 &  0.32 &  0.06 &  0.27 &  0.07 \\
    6304 &  1.14 &  0.01 & -0.61 &  0.02 & -0.11 &  0.05 &  0.46 &  0.05 &  0.31 &  0.02 &  0.20 &  0.04 &  0.17 &  0.04 &  0.30 &  0.04 \\
    6316 &  1.14 &  0.01 & -0.87 &  0.03 & -0.11 &  0.06 &  0.36 &  0.08 &  0.45 &  0.03 &  0.14 &  0.07 &  0.32 &  0.05 &  0.37 &  0.06 \\
    6333 &  0.89 &  0.01 & -1.73 &  0.03 &  0.16 &  0.05 & -0.08 &  0.12 &  0.43 &  0.07 &  0.32 &  0.12 &  0.40 &  0.08 &  0.20 &  0.05 \\
    6342 &  1.06 &  0.05 & -0.89 &  0.05 & -0.03 &  0.09 &  0.44 &  0.15 &  0.45 &  0.06 &  0.09 &  0.13 &  0.19 &  0.11 &  0.29 &  0.11 \\
    6352 &  1.14 &  0.01 & -0.72 &  0.02 & -0.05 &  0.02 &  0.29 &  0.03 &  0.45 &  0.02 &  0.26 &  0.03 &  0.21 &  0.03 &  0.37 &  0.03 \\
    6356 &  1.14 &  0.01 & -0.75 &  0.01 &  0.09 &  0.03 &  0.55 &  0.04 &  0.41 &  0.02 &  0.25 &  0.03 &  0.22 &  0.03 &  0.38 &  0.03 \\
    6362 &  1.05 &  0.03 & -1.30 &  0.03 &  0.18 &  0.05 &  0.20 &  0.13 &  0.70 &  0.04 &  0.36 &  0.09 &  0.34 &  0.07 &  0.46 &  0.07 \\
    6388 &  0.94 &  0.01 & -0.63 &  0.01 & -0.27 &  0.01 &  0.44 &  0.02 &  0.10 &  0.01 &  0.15 &  0.02 &  0.08 &  0.02 &  0.20 &  0.02 \\
    6441 &  0.95 &  0.01 & -0.58 &  0.01 & -0.27 &  0.01 &  0.42 &  0.02 &  0.24 &  0.01 &  0.18 &  0.02 &  0.13 &  0.02 &  0.23 &  0.02 \\
    6522 &  0.86 &  0.02 & -1.21 &  0.03 & -0.11 &  0.06 &  0.56 &  0.09 &  0.47 &  0.03 &  0.30 &  0.07 &  0.35 &  0.05 &  0.40 &  0.06 \\
    6528 &  1.07 &  0.03 & -0.23 &  0.02 &  0.09 &  0.02 &  0.43 &  0.04 &  0.19 &  0.02 &  0.17 &  0.03 &  0.03 &  0.04 &  0.09 &  0.04 \\
    6544 &  1.14 &  0.01 & -1.36 &  0.04 & -0.09 &  0.05 &  0.11 &  0.20 &  0.16 &  0.04 &  0.08 &  0.07 &  0.29 &  0.09 &  0.47 &  0.08 \\
    6553 &  0.89 &  0.02 & -0.23 &  0.02 & -0.14 &  0.03 &  0.50 &  0.03 &  0.29 &  0.01 &  0.13 &  0.02 & -0.01 &  0.02 &  0.21 &  0.03 \\
    6569 &  1.04 &  0.04 & -1.07 &  0.04 &  0.11 &  0.05 &  0.53 &  0.10 &  0.53 &  0.05 &  0.29 &  0.10 &  0.29 &  0.09 &  0.45 &  0.09 \\
    6624 &  1.12 &  0.02 & -0.77 &  0.01 & -0.15 &  0.02 &  0.42 &  0.03 &  0.36 &  0.01 &  0.25 &  0.03 &  0.23 &  0.02 &  0.41 &  0.03 \\
    6626 &  0.83 &  0.01 & -1.27 &  0.02 &  0.05 &  0.03 &  0.68 &  0.06 &  0.57 &  0.03 &  0.32 &  0.06 &  0.33 &  0.05 &  0.40 &  0.05 \\
    6637 &  1.14 &  0.01 & -0.90 &  0.01 &  0.16 &  0.01 &  0.54 &  0.03 &  0.43 &  0.02 &  0.20 &  0.03 &  0.30 &  0.03 &  0.54 &  0.02 \\
    6638 &  1.02 &  0.02 & -1.02 &  0.02 &  0.12 &  0.05 &  0.64 &  0.07 &  0.41 &  0.03 &  0.24 &  0.06 &  0.32 &  0.05 &  0.52 &  0.05 \\
    6652 &  1.01 &  0.01 & -0.93 &  0.02 &  0.02 &  0.02 &  0.23 &  0.05 &  0.45 &  0.02 &  0.23 &  0.04 &  0.25 &  0.03 &  0.33 &  0.03 \\
    6723 &  1.00 &  0.02 & -1.32 &  0.03 & -0.00 &  0.04 &  0.27 &  0.11 &  0.53 &  0.04 &  0.20 &  0.10 &  0.33 &  0.06 &  0.29 &  0.06 \\
    6752 &  0.87 &  0.01 & -1.64 &  0.03 &  0.16 &  0.04 &  0.02 &  0.14 &  0.54 &  0.05 &  0.42 &  0.09 &  0.33 &  0.07 &  0.29 &  0.06 \\
    7078 &  0.86 &  0.01 & -2.09 &  0.01 &  0.29 &  0.03 &  0.43 &  0.25 &  0.67 &  0.08 &  0.54 &  0.14 &  0.35 &  0.05 &  0.32 &  0.01 \\
    7089 &  0.86 &  0.01 & -1.54 &  0.02 &  0.12 &  0.05 & -0.12 &  0.12 &  0.45 &  0.04 &  0.36 &  0.07 &  0.28 &  0.07 &  0.24 &  0.05
\enddata
\vspace{0.1cm} 
\tablecomments{Ages are quoted in logarithmic units.  Errors represent
formal $1\sigma$ statistical uncertainties only.}
\label{t:res}
\end{deluxetable*}

\subsection{Comparison of Spectral and Index Fitting}
\label{s:spec_vs_indx}

\subsubsection{Overview}
\label{s:indxoverview}

Our choice to fit the absorption line spectrum directly, rather than
measuring and fitting in index space, is in contrast to nearly all
other modeling approaches \citep[e.g.,][]{Trager00a, Thomas05,
  Schiavon07, Graves08, Thomas11b}.  There are many advantages to
fitting spectra directly.  One important feature is the ability to
carefully inspect model residuals to identify regions where either the
data or the model are poor (comparing residuals in the restframe and
observed frame allows one to distinguish between these two
possibilities).  Relatedly, bad regions in the data (e.g., where sky
line residuals are strong) can be easily masked when fitting the
spectrum directly.

Another key benefit of fitting spectra rather than indices is the
higher information content of the former.  As first demonstrated in
\citet[][their Fig. 7]{Sanchez-Blazquez11}, for a fixed S/N spectrum,
fitting in index space results in substantially larger uncertainties
and degeneracy between age and metallicity than when fitting the
spectrum directly.  This should not be surprising, since one is
effectively taking $\sim1000$ independent data points in a spectrum
and reducing them to $\sim10-20$ data points (although these $10-20$
data points were specifically chosen to be information-dense, so
the loss is not as substantial as the raw numbers would suggest).
Moreover, those small number of data points represent a substantial
effective smoothing of the spectrum due to the fact that the indices
are measured over $10-50$\AA\, windows.  This smoothing both reduces
the overall information content and introduces degeneracies between
parameters because numerous spectral features that were once clearly
separated in spectral space now combine to contribute to a single
index \citep[as an example, the CN bands contribute to the blue
pseudo-continuum of the H$\delta$ Lick index, greatly complicating the
interpretation of that index;][]{Prochaska07, Schiavon07}.

There is, in our view, one notable advantage of working with indices
rather than the full spectrum and that is the diagnostic power and
conceptual simplicity of considering a small number of features.
Index-index diagrams can be powerful tools for identifying
degeneracies between parameters and isolating primary sensitivities of
particular features.  They can also be useful for visualizing trends
in the data.  These benefits are most obvious when considering the
primary parameters of age, metallicity, and $\alpha$-abundance.

\subsubsection{Tests with mock data}

In order to quantitatively explore the comparison between fitting
indices and spectra, we have fit a set of mock data using both
approaches.  The mock data were computed with an age of 10 Gyr,
solar-scaled abundances, a velocity dispersion of $150\kms$, and at
two metallicities, [Fe/H]=+0.0 and -1.0.  We generated mock spectra
with S/N=20, 30, 50, and 100 per \AA.  For each S/N and metallicity,
the full spectrum from $0.40-0.56\mu m$ was fit with \alf, and then we
measured 14 spectral indices from the same spectrum, and fit those 14
indices with \alf.  In both cases the underlying fitting model is the
same - we fit with \alf\, in simple mode as described above.  We chose
to fit the following indices: H$\delta_F$, CN$_2$, Ca4227, G4300,
H$\gamma_F$, Fe4383, Fe4531, C$_2$4668, H$\beta$, Fe5015, Mg{\it b},
Fe5270, Fe5335, Fe5406.  This is the set of Lick indices usually
considered in the literature \citep[e.g.,][]{Graves08, Thomas11b,
  Johansson12}, although see \citet{Serven05} who define a much larger
set of indices in an attempt to measure many more elements within the
index-based framework\footnote{We note that it is fairly common in the
  literature to include multiple indices with strongly overlapping
  wavelength ranges. An example of this is CN$_1$ and CN$_2$, which
  only differ only in the placement of the blue pseudo-continuum.  It
  is not appropriate to include both indices in the fitting unless the
  (very strong) covariance between the indices is included in the
  likelihood calculation.}.

The results are shown in Figures \ref{fig:indzp} and \ref{fig:indzm}.
In each panel we show the symmetrized $1\sigma$ errors, (84\% CI-16\%
CI)/2., as a function of S/N.  Figure \ref{fig:indzp} shows mock data
at [Fe/H]=+0.0 and Figure \ref{fig:indzm} shows mock data at
[Fe/H]=-1.0.  In each panel, results from fitting the full spectrum
(solid lines) are compared to results from fitting indices (dashed
lines).  There are several important conclusions to draw from these
figures.  First, in every case the spectral fitting outperforms the
indices, in many cases by large margins.  At low metallicity and
moderately low S/N ($\le 50$), the errors on [Fe/H] are larger by
$>0.1$ dex when fitting indices.  Phrased another way, the errors on
[Fe/H] and many other elements from fitting spectra at S/N=20 are
comparable to the errors from fitting indices at S/N=50.  To achieve
the same precision when fitting indices would therefore require
$(50/20)^2=6.25$ times longer integration compared to fitting spectra.

It is also worth pointing out that at S/N=100 and [Fe/H]$=+0.0$ the
spectral fitting results return errors less than 0.05 dex for all
parameters and therefore, at least for these parameters, even higher
S/N is unlikely to be helpful because of uncertainties in the
abundance scale of the empirical library \citep[but see][for examples
of parameters that require S/N$>300$]{Conroy17a}.  Even at
[Fe/H]$=-1.0$ nearly all parameters are measured to a precision of
$<0.1$ dex at S/N=100.

Given that the allowed range of the parameters only extends from -0.3
to +0.5 for generic [X/H] (see Table \ref{t:alf}) there is a limit to
the allowed error range (the limit cannot be directly inferred from
the ranges listed in Table \ref{t:alf} because in Figures
\ref{fig:indzp} and \ref{fig:indzm} we are quoting results in [X/Fe] =
[X/H] - [Fe/H]) and it appears that the errors on [Ti/Fe] in Figure
\ref{fig:indzm} are running up against that limit and so the errors in
these cases are likely underestimated.

Finally, in Fig \ref{fig:corner} we show the posterior probability
distributions for age, [Fe/H], and [Mg/Fe] for a mock dataset with
S/N$=30$.  We compare full spectrum fitting with index-based
constraints.  Not only are the constraints stronger when fitting the
full spectrum but also the degeneracy between these three parameters
is smaller compared to indices.

\subsubsection{Summary}
\label{s:indxsum}

The conclusion we draw from the tests in the previous section is that
fitting observations directly in spectral space offers more precise
measurements of many parameters, which translates directly into less
demanding S/N and exposure time requirements \citep[see also][their
Fig. 7]{Sanchez-Blazquez11}.  We have not undertaken an exhaustive
study, and other, more expansive index sets \citep[e.g.,][]{Serven05}
may reduce the discrepancy between index and full spectrum fitting
uncertainties shown above.  Indeed, in the limit of many
non-overlapping indices, the methodology becomes similar except for a
significant effective smoothing that occurs in the index analysis. As
noted in Section \ref{s:indxoverview}, there are other benefits of
full spectrum fitting including inspecting residuals, straightforward
masking of bad data, and avoiding correlated errors that come with
indices that have overlapping wavelength boundaries.  Tests of the
effect of imperfect flux calibration on modeling spectra and indices
are presented in Appendix B.

The mock tests from the previous section can also be used to dispel a
common misconception regarding full spectrum fitting --- that
parameters which affect only one or a few isolated wavelength regions
(e.g., Ba or Sr) are better modeled with indices because fitting the
full spectrum ``dilutes'' the signal in such cases.  In reality, the
addition of wavelength regions that have no effect on the model
parameter of interest will in turn have no effect on the constraints
for that parameter.  In other words, there is no dilution.  The mock
tests confirm this - if dilution of the signal were occurring then we
would expect indices to perform as well or better for those elements
that have fairly localized effects on the spectrum, e.g., Mg, Ca, C,
and N in the example above.  However, the opposite behavior is seen in
the mock tests.  We explore this issue further in Appendix A, where we
demonstrate with a controlled test that the addition of larger and
larger wavelength intervals does not impact the derived parameter
values, even in the case where larger wavelength intervals bring in
more systematic residuals.

Nonetheless, we emphasize that full spectrum fitting is not a panacea.
A fitting technique is only as good as in the input models and data,
and there are clear systematic issues at the $0.05-0.1$ dex level, as
we demonstrate below when fitting to globular cluster data, and as
also shown from comparison of several state-of-the-art fitting
techniques when applied to early-type galaxy data in
\citet{Conroy14a}.  In the limit of systematics-limited models one must
exercise an abundance of caution with any fitting algorithm,
especially a many dimensional problem in which multiple parameters may
compensate for a particular model deficiency.  At the end of the
day, the best test of any model is the comparison to external
calibration samples, which is why we now turn to a comparison with
Galactic globular clusters.

\begin{figure*}[!t]
\center
\includegraphics[width=0.9\textwidth]{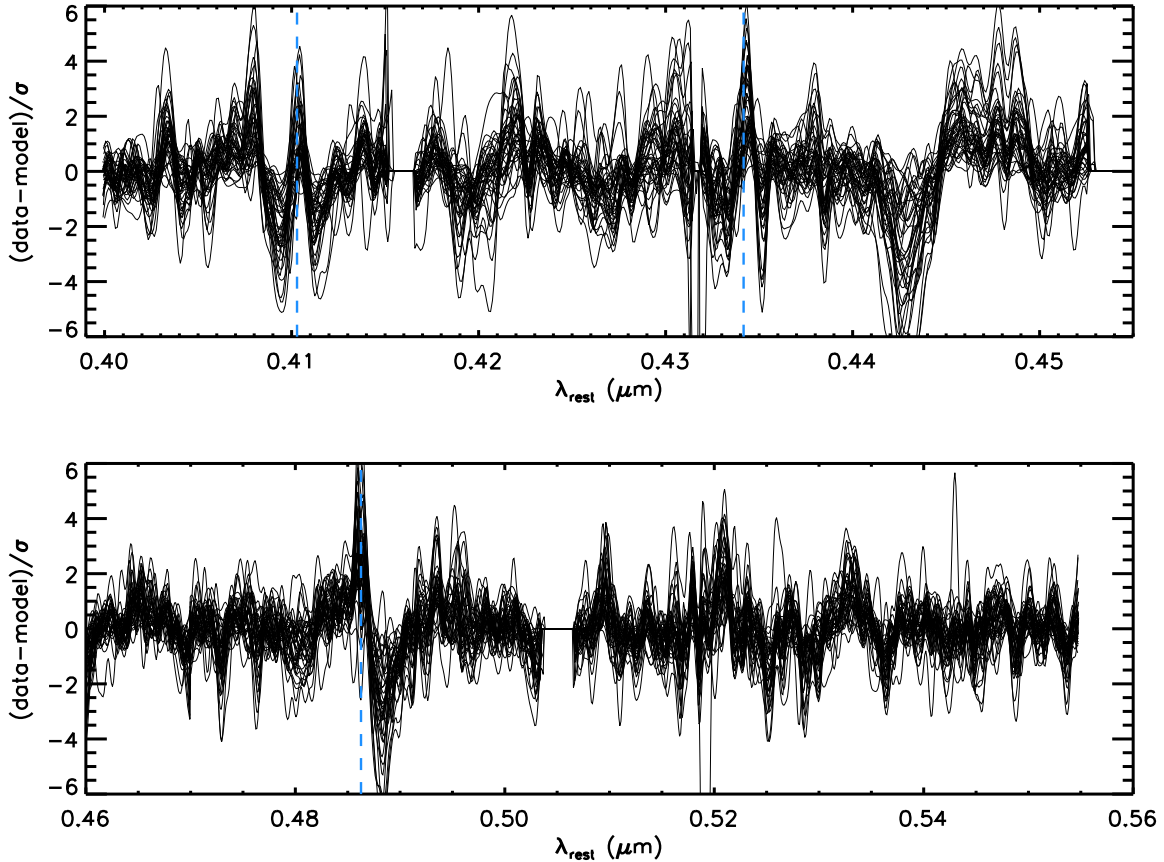}
\caption{Residuals of the fits to all 41 Galactic GCs, plotted as
  data-model divided by the errors on the data.  Spectra have been
  shifted to the restframe.  Vertical dashed blue lines mark the
  wavelengths of the H$\delta$, H$\gamma$ and H$\beta$ lines. }
\label{fig:s05resid}
\end{figure*}


\section{Fits to Galactic Globular Clusters}
\label{s:res}

\subsection{Setup}
\label{s:s05}

We now turn to fitting integrated light spectra of Galactic GCs in
order to validate the new models described above.  The value of
testing models against GCs is that their ages, metallicities, and
detailed abundance patterns can be derived from the color-magnitude
diagram and high resolution spectroscopy of individual stars.  The use
of GCs as calibrators is not without its challenges, a point we return
to in Section \ref{s:disc}.

We make use of the library of integrated spectra of Galactic GCs
presented in \citet[][S05]{Schiavon05}.  These authors obtained
optical spectra of 41 GCs covering the range $0.33-0.64\mu m$ at a
resolution of $\approx3.1$\AA\, (FWHM).  The physical extent of GCs
makes it challenging to obtain true integrated spectra of them.  S05
were able to sample the cluster core diameter by letting the cluster
drift across the slit during the 15m long exposure.  Due to the
extended nature of GCs, the fact that the brighest stars are resolved,
and the crowded fields of many clusters (especially those in the
bulge), the extraction of reliable high quality spectra is extremely
challenging and we refer the interested reader to S05 for details.

From initial inspections and fits we identified several wavelength
intervals that appeared to be corrupted and so we masked them from the
final fitting.  These regions are: $4150-4165$\AA, $4313-4318$\AA,
$4525-4565$\AA, and $5033-5065$\AA.  We avoid the redder segment of
the S05 data as the region around $0.59\mu m$ is strongly contaminated
in many spectra via NaD absorption associated with the interstellar
medium.  S05 quote a wavelength-dependent instrumental resolution for
their spectra, which we use to smooth all of our input models at the
initial setup stage.  Owing to the fact that the intrinsic resolution
of Galactic GCs is higher than our input empirical models (which are
smoothed to a common resolution of $\sigma=100\kms$), we are forced to
smooth the S05 spectra.  We chose to smooth the S05 spectra by a
Gaussian with $\sigma=150\kms$.  We emphasize that modest changes in
the spectral resolution in the range of $\sigma=100-350\kms$ does not
alter the recoverability of parameters, as demonstrated in
\citet{Choi14}.  Finally, the S05 spectra were converted from air to
vacuum wavelengths as our models are on a vacuum wavelength scale.  We
fit these 41 spectra over the wavelength range $0.40-0.555\mu m$ with
\alf\, in simple mode.

\begin{figure*}[!t]
\center
\includegraphics[width=0.96\textwidth]{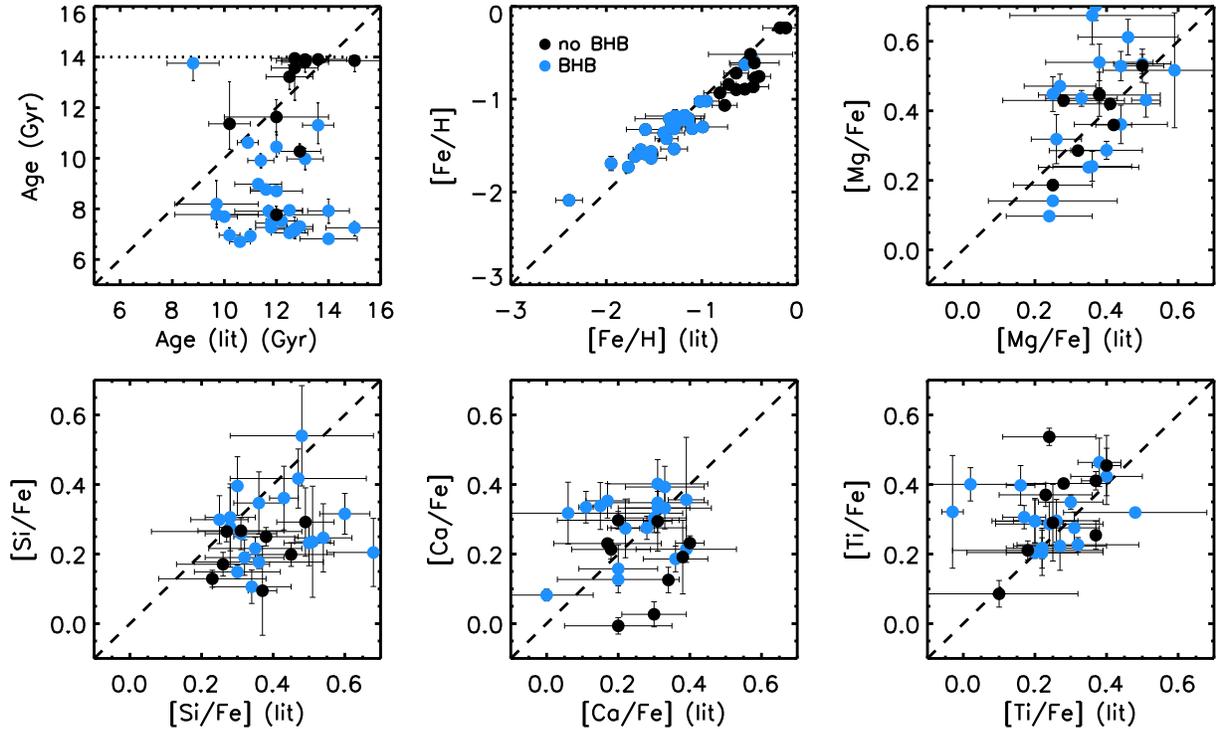}
\vspace{0.1cm}
\caption{Comparison of parameters derived by fitting the integrated
  light spectra of Galactic GCs to literature values provided in the
  compilation of \citet{Roediger14}.  Black points are GCs that lack a
  prominent blue horizontal branch (BHB) while blue points have a
  prominent BHB.  The dotted line at 14 Gyr marks the upper limit of
  our allowed range in age.}
\label{fig:resbhb}
\end{figure*}

\subsection{Results}

We begin with Figure \ref{fig:s05ex} by comparing the best-fit models
to the spectra for three GCs spanning a range in metallicity.  The
upper panel for each GC shows the continuum-normalized spectra and the
lower panel shows the ratio between model and data (in the sense of
data/model) compared to the errors on the observations (shaded band).
It is encouraging that the residuals are nearly always encompassed
within the $1\sigma$ errors of the data.

Residuals for all 41 spectra are shown in Figure \ref{fig:s05resid},
where now the residuals are shown in units of the errors in the data.
Two important features stand out.  First, the overall level of the
residuals is small and generally consistent with the observational
errors (the median of $|$data-model$|/\sigma$ is 1.00).  Second, there
are strong correlations in the residuals across the population.  Some
of the features correspond to obvious features in the data and/or
models, such as H$\beta$ at $0.486\mu m$.  The broad feature at
$0.443\mu m$ is interesting.  It is not associated with an obvious,
strong feature in the GC spectra.  We computed the rms residual around
this feature and searched for correlations between the rms and derived
GC parameters.  The rms correlates strongly with both [Fe/H] and
[C/Fe], though the correlation is stronger with the latter parameter.
Most strikingly, the two GCs with the largest rms, NGC 6388 and NGC
6441, are also the two GCs with the {\it lowest} [C/Fe] abundance
([C/Fe]=-0.27).  We discuss a possible explanation for this behavior
in Section \ref{s:disc}.

Table \ref{t:res} presents the best-fit parameters for the 41 Galactic
GCs from S05.  We do not include [O/Fe] as it appears to be poorly
constrained, which is not surprising given that we are only fitting
the $0.4-0.555 \mu m$ wavelength range, where there are few unique
Oxygen-related features at low resolution (see Figure \ref{fig:cno}).

\begin{figure*}[!t]
\center
\includegraphics[width=0.96\textwidth]{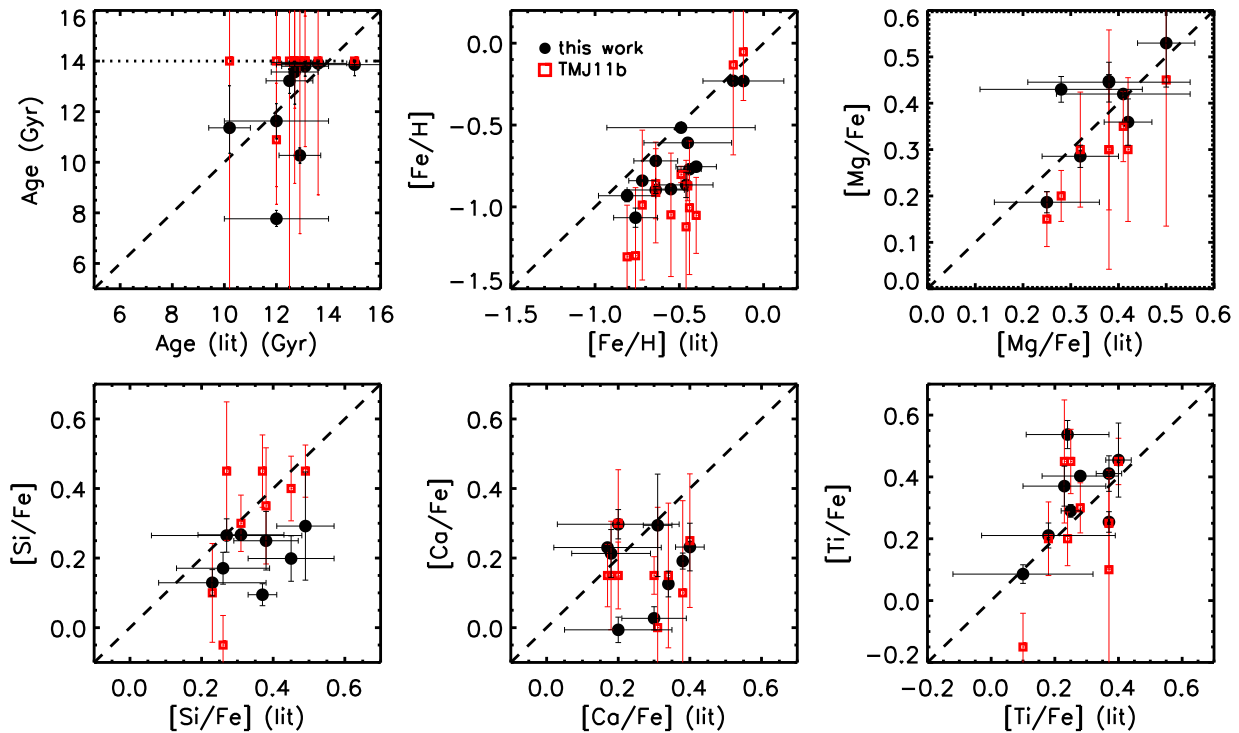}
\vspace{0.1cm}
\caption{As in Figure \ref{fig:resbhb}, now restricting the sample to
  GCs without a prominent BHB.  We also include results from
  \citet[][TMJ11b]{Thomas11b} who fit the same data using Lick
  indices.  The TMJ11b results allow ages $>14$ Gyr; for consistency
  with our results we have plotted their results at 14 Gyr if the
  best-fit age exceeds 14 Gyr.  We have omitted x-axis errors on the
  TMJ11b results for clarity.}
\label{fig:resnobhb}
\vspace{0.1cm}
\end{figure*}

The derived parameters from the integrated light spectra are compared
to literature values in Figures \ref{fig:resbhb} and
\ref{fig:resnobhb}.  Literature values are adopted from the
compilation presented in \citet{Roediger14}.  Note that the error bars
are quite large as they encompass both systematic errors revealed by
different abundances reported by different authors and true intrinsic
spreads of some elements in at least some GCs; see \citet{Roediger14}
for details.

In Figure \ref{fig:resbhb} we color-code points by whether or not they
contain a prominent blue horizontal branch (BHB) population.  These
are mostly identified by the HBR parameter in the \citet{Harris96}
catalog, with additional identification directly from inspection of
available CMDs.  As is well-known, GCs with prominent BHB populations
are challenging to model because the additional hot stars, rarely
included in models, results in a spectrum that appears much younger
than reality \citep[e.g.,][]{Worthey94, Schiavon07}.  We find the same
result in Figure \ref{fig:resbhb} - GCs with prominent BHB populations
have derived ages considerably younger than the literature values.
The metallicities are nonetheless accurately recovered, even when
strong BHB populations are present.  This result can be understood
through the behavior of the age-metallicity vectors as e.g., discussed
in \citet{Worthey94}.  At low metallicity the vectors are nearly
orthogonal in the sense that Fe features are nearly insensitive to
age.  However, the other elements shown in Figure \ref{fig:resbhb},
Mg, Si, Ca and Ti, show considerable scatter for those clusters with
prominent BHB populations.

In Figure \ref{fig:resnobhb} we focus on GCs lacking a prominent BHB
population.  We also include a comparison to the results from
\citet[][TMJ11b]{Thomas11b} who fit the same S05 spectra using their
Lick index-based models.  The overall agreement between the
literature, our results, and those of TMJ11b is encouraging.  We
derive slightly higher [Fe/H] metallicities than TMJ11b, in closer
agreement with the literature.  Both our and TMJ11b's ages tend to
exceed the upper limit imposed by the age of the Universe, a common
problem for all models.  The Mg and Ti abundances agree very well with
the literature.  Si and Ca in contrast show less agreement, although
the dynamic range in smaller.  The reported errors in TMJ11b are much
larger than reported here.  It is beyond the scope of this paper to
explore this in detail, though we note that the generally larger
errors associated with Lick index fitting compared to full spectrum
fitting is in qualitative agreement with the mock tests reported in
Section \ref{s:spec_vs_indx}.


\section{Discussion}
\label{s:disc}

\subsection{Caveats and limitations}

Stellar population models bring together many ingredients including
empirical stellar libraries, isochrones, theoretical response
functions, an assumed parameterization (e.g., the star formation
history, grouping certain elements together, fixing or fitting for the
IMF), and fitting techniques.  Each of these pieces are active areas
of development and none are free of systematics \citep[see][for a
review]{Conroy13b}.  The theoretical spectral models contain many
well-documented limitations including incomplete line lists, adopted
microturbulence, assumptions of 1D hydrostatic atmospheres and LTE,
etc.  \citet{Martins07, Sansom13} has presented comparisons between
models and empirical spectra and find some encouraging results but
also significant limitations and areas for improvement.  The empirical
libraries do not uniformly sample all relevant regions of parameter
space (including $\teff$, $\logg$, [Z/H], and abundance patterns), and
are often limited by various data reduction issues including proper
absolute flux calibration and telluric correction \citep[see][for
discussion]{Villaume17a}.  And as we have discussed in Section
\ref{s:spec_vs_indx}, the choice of fitting technique can affect the
precision of the derived parameters, even in the limit of perfect
models.

There is a robust community addressing every one of the limitations
noted above, and progress continues at a steady pace.  Nonetheless it
can be difficult to understand the integrated effect of all these
limitations on the derived parameters.  The only robust end-to-end
test that we are aware of is testing the models against well-studied
calibrators such as globular clusters, although these systems may not
be ideal calibrators for reasons discussed below.  Identifying
additional, independent tests of the models would be a major advance.

\begin{figure*}[!t]
\center
\includegraphics[width=0.95\textwidth]{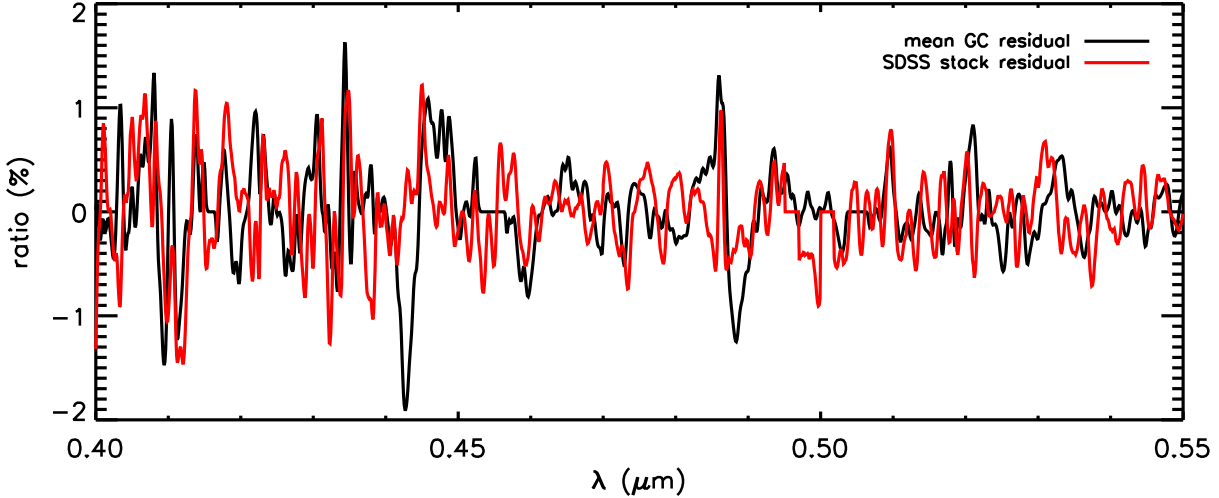}
\vspace{0.1cm}
\caption{Comparison of the mean ratio between data and model from
  fitting the Galactic GCs (black line) and fitting a stacked spectrum
  of early-type galaxies from SDSS (red line).  All of the spectra
  have been smoothed to the same resolution ($\sigma=150\kms$ plus an
  instrumental resolution of $\approx80\kms$).  The residual levels
  are broadly comparable except at $0.443\mu m$ and $0.488\mu m$,
  where the GC residuals are much larger than the SDSS residuals.  The
  origin of these residuals is unclear but they may be connected to
  the chemical peculiarity of GCs. }
\label{fig:rescomp}
\end{figure*}

\subsection{GCs as calibrators of stellar population models}

We have shown that the new models presented herein perform slightly
better than previous models when comparing to GCs, but discrepancies
remain. This could mean that our models are incomplete (as noted
above), but these issues also raise the question whether GCs are
actually a suitable calibration for the kind of detailed modeling we
are performing here.

GCs have long been used as a benchmark calibration sample for stellar
population models \citep[e.g.,][]{Worthey94, Schiavon07, Thomas11b,
  Conroy14a}.  They are valued because resolved star CMDs can deliver
precise ages and high resolution spectra of individual stars can be
used to measure precise abundances of many elements.  They were also
long believed to be mono-metallic and mono-age, which if true would
make them genuine empirical SSPs.  However, the use of GCs as
calibrators is not without challenges.  It is now well known that most
if not all GCs harbor substantial internal abundance spreads in at
least the light elements C, N, O, Na, Mg, and Al, and the more massive
clusters also show evidence for variation in He, Ca, and Fe \citep[see
review in][]{Gratton04}.  The most extreme example of this phenomenon
is $\omega$ Cen, which shows a variation in [Fe/H] of $\approx0.3$ dex
and a He spread of $\Delta {\rm He}\approx0.14$ \citep{Piotto05}.  The
variation in [O/Fe] and [Na/Fe] exceeds 1 dex \citep{Johnson10} and
the CMD shows an incredibly rich and complex structure
\citep{Bedin04}.  Whatever physical process is giving rise to this
complexity seems to be isolated to the GC environment - field stars in
the Galaxy do not show such extreme abundance variations
\citep{Gratton04}.

In addition to the internal complexity of many GCs, other factors
diminish their utility as calibrators.  GCs are subject to (modest)
stochastic effects due to the limited number of very luminous giants
\citep{Cervino04} and this effect on the line strengths has not been
extensively studied.  As discussed in Section \ref{s:s05}, it is
challenging to obtain integrated-light spectra of GCs in the Galaxy
due to their extended nature and the fact that some, especially the
metal-rich bulge clusters, are in crowded regions.  As a practical
matter existing GC integrated light spectra had been limited to the
optical, although this is being remedied with the WAGGS project which
is collecting IFU spectra of GCs from $0.33-0.95\mu $
\citep{Usher17}.  Finally, as already discussed in Section
\ref{s:res}, many GCs contain large populations of BHB stars, which
  compromise age estimates from integrated light spectra.  Much of the
  BHB phenomenon at higher metallicities is likely linked to the
  chemically peculiar nature of GCs, which suggests that the
  phenomenon may be localized to GCs. 

In spite of these myriad challenges, we have shown that our models are
able to recover the overall metallicity for all 41 GCs in the S05
sample, and for GCs lacking a prominent BHB population, we are also
able to recover the abundances of Mg, Si, Ca, and Ti.  The residuals
between the data and best-fit model also appear well-behaved, which is
an important test.  However, in Figure \ref{fig:s05resid} there were
clear correlations in the residuals across the sample, and as we
discussed in Section \ref{s:res}, at least one strong residual at
$0.443\mu m$ appears to correlate strongly with the derived [C/Fe]
abundance.  The residuals appear larger when [C/Fe] is lower, which
suggests a correlation with GC chemical peculiarity (more chemically
extreme GCs show lower average [C/Fe] abundances).

In order to explore this further we compare the mean GC residuals to
the residual from fitting a stacked spectrum of SDSS early-type
galaxies in Figure \ref{fig:rescomp}.   The SDSS spectrum is from the
sample described in \citet{Conroy14a}.  Specifically we selected
galaxies in a narrow velocity dispersion bin, $125<\sigma<150\kms$ and
smoothed all of the spectra to the highest dispersion in the bin.  We
fit the same spectral range as the GC data and with the same \alf\,
setup.  The \ionn{O}{iii]} lines were masked.  It is encouraging that
the overall residual level is comparable given that completely
different datasets and sources.  The only two features that stand out
are the $0.443\mu m$ and $0.488\mu m$ residuals in the GC data that do
not show up strongly as residuals in the SDSS galaxy data.    One
possibility is that the S05 spectra are somehow corrupted in these
wavelength regions, although we think this explanation unlikely
because these residuals are in the restframe and they correlate with
[C/Fe], which is measured from spectral features at other wavelengths.
An intriguing possibility is that these residuals are a marker of the
unique chemical complexity of GCs.  We leave the exploration of this
possibility to future work.

\subsection{Future directions}

In our approach to modeling absorption line spectra we have removed
the continuum information by multiplying the models by polynomials.
In ongoing work we are exploring ways to combine the information in
the narrow spectral features with the overall continuum shape
\citep[specifically broadband photometry; see e.g.,][]{Gu17}.
Especially for low S/N spectra the broadband optical colors appears to
provide important independent constraints (Johnson et al. in prep.).

There are several improvements to the models that we are actively
pursuing.  We are observing additional MILES stars with the IRTF in
order to expand and fill out the low metallicity and younger age
regions of parameter space.  This should improve the reliability at
the lowest metallicities and expand the age coverage down to 100 Myr.
We are finalizing an empirical calibration of the atomic and molecular
line lists (Cargile et al. in prep) that should result in more
accurate synthetic models and therefore more accurate response
functions.

We are also computing new isochrone tables as part of the MIST project
\citep{Dotter16, Choi16}.  These new isochrones will include not only
overall metallicity variation but also [$\alpha$/Fe] variation, and
some selected models with individual element variation.  These will
eventually be incorporated into the empirical and theoretical SSPs.


\section{Summary}

We have presented a new set of stellar population models appropriate
for modeling old ($>1$ Gyr) systems spanning a wide range in
metallicity ($-1.5\lesssim$ [Fe/H] $\lesssim0.3$) and covering a wide
wavelength range ($0.37-2.4 \mu m$).  These models include the ability
to independently vary the abundances of 18 elements, as well as a
variety of other effects.  We used these new models to fit integrated
light spectra of Galactic GCs from S05.  We now summarize our main
results:

\begin{itemize}

\item  These new models compare well with existing models from S07,
  V10, and TMJ11 in terms of the age and metallicity-dependence of the
  classic Lick indices.

\item We present new metallicity and age-dependent element response
  functions covering the full optical and NIR wavelength range.  As
  expected, the sensitivity to element variation decreases both for
  younger ages and lower metallicities.

\item We present NLTE models of several strong Na and Ca lines and
  conclude that the assumption of LTE in the construction of the
  theoretical response functions is adequate at the sub-percent
  level.

\item Model fits to the optical spectra of Galactic GC spectra
  reproduce the data at the $\lesssim 1$\% level, with typical
  residuals within $1\sigma$ of the observations.  The derived
  metallicities, [Fe/H], agree well with literature values.  For GCs
  lacking a prominent BHB population, we are also able to recover the
  [Mg/Fe], [Si/Fe], and [Ti/Fe] abundances, while [Ca/Fe] abundances
  are somewhat less-well recovered.

\end{itemize}


\acknowledgments 

We thank Russell Smith for providing tabulated atmospheric absorption
models, Dan Foreman-Mackey for porting \texttt{emcee} to fortran, and
the referees for comments that sharpened the arguments in this paper.
C.C. acknowledges support from NASA grant NNX15AK14G, NSF grant
AST-1313280, and the Packard Foundation.  A.V. acknowledges the
support of an NSF Graduate Research Fellowship.


\begin{appendix}

\section{Narrow features in the context of full spectrum fitting: dilution or not?}

It is clear from Figures \ref{fig:cno} and \ref{fig:femgti} \citep[see
also Figures 3-4 in][]{Conroy14a} that some elements affect only a
narrow wavelength range.  One may then wonder if modeling the full
spectrum is a sensible approach for such parameters, or whether the
feature is ``diluted'' in some way by simultaneously considering a
wide wavelength range.  This is an important question in the context
of fitting the full spectrum or line indices; for elements that
produce many lines at many wavelengths (such as Fe) it is self-evident
that a full spectrum approach is preferable, but for elements whose
main signature is limited to a narrow wavelength interval (such as N,
Ca, Mn, Sr, Ba, or Eu) one might reason that limiting the fit to that
particular wavelength region would give the most reliable results.

We have already discussed this point in Section \ref{s:indxsum} in the
context of fits to mock data.  Here we consider this issue again with
a controlled experiment.  We have created an artificial dataset in
which one line of interest at $0.5\mu m$ is modeled in the presence of
numerous additional unmodeled lines.  Each line is a Gaussian.  The 20
unmodeled lines are given random line parameters within the wavelength
range $0.3-0.7\mu m$.  The adopted S/N of the dataset is 50 per pixel.
We then model the line depth and width of the $0.5\mu m$ line for
three wavelength intervals: narrow (200\AA), medium (1000\AA) and wide
(3000\AA) settings.  The procedure is similar to \alf\, in that we use
MCMC techniques with the likelihood in Equation 2 to derive the
posterior constraints.

\begin{figure*}[!t]
\center
\includegraphics[width=0.9\textwidth]{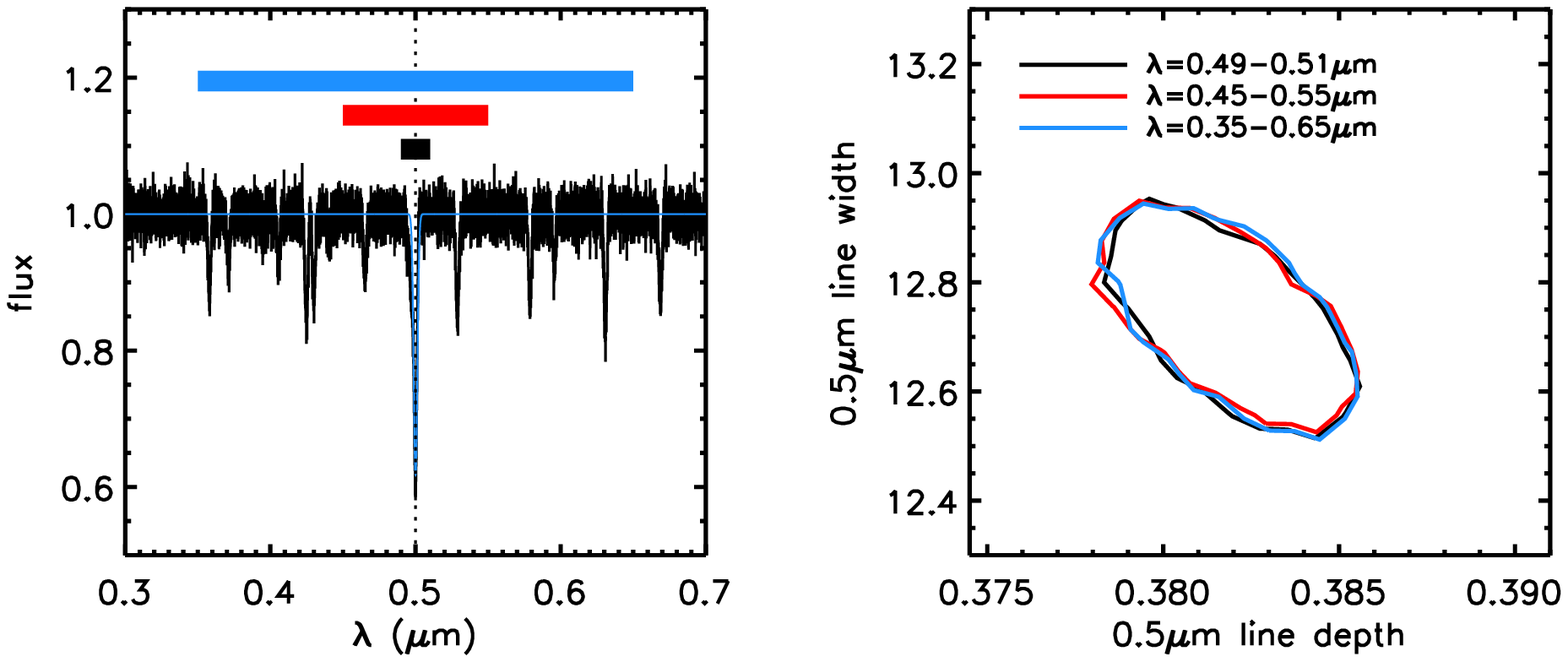}
\vspace{0.1cm}
\caption{Effect of wavelength interval on measuring the parameters of
  a narrow absorption feature.  An artificial dataset is constructed
  in which one Gaussian line at 0.5$\mu m$ is modeled in the presence
  of both noise and additional unmodeled lines at random wavelengths
  and line depths.  {\it Left panel:} The dataset is shown as a black
  line and the horizontal bands show the three fitted wavelength
  intervals.  In each case the line depth and width of the 0.5$\mu m$
  line were fit.  The thin blue line shows the best fit model for the
  widest wavelength setting.  {\it Right panel:} 68\% CI of the fitted
  0.5$\mu m$ line depth and width for three wavelength intervals
  including narrow (200\AA), medium (1000\AA) and wide (3000\AA)
  settings.  The constraints are identical, demonstrating that fitting
  over a wide wavelength range that contains model deficiencies
  does not `dilute' the constraints inferred for the narrow feature of
  interest.}
\label{fig:dilution}
\end{figure*}

The result of this experiment is shown in Figure \ref{fig:dilution}.
In the left panel we show the dataset, and we note that by chance one
of the random, unmodeled lines falls just to the left of the
$0.5\mu m$ line.  In the right panel we show the 68\% CI joint
constraints on the depth and width of the 0.5$\mu m$ line.  Results
are shown for the three wavelength intervals and it is clear that the
results are identical.

The inclusion of additional wavelengths therefore does not deteriorate
the fit to a narrow feature, even when the additional wavelengths
include systematic model uncertainties (in our case unmodeled lines).
Of course, the minimum $\chi^2$ values are quite different in each
case, ranging from 2156 to 29922.  But it is important to remember
that parameter confidence intervals are determined not by the absolute
values of $\chi^2$ but rather by {\it differences} in $\chi^2$.  The
addition of uninformative wavelength intervals adds a constant
pedestal to $\chi^2$, but does not impact the parameter constraints.

\section{The impact of imperfect fluxing on parameter recovery}

Even in the best cases the flux calibration of spectra over a wide
wavelength range is rarely better than $5-10$\%, and often it is much
worse.  The factors that must be accounted for include the instrument
response function, the detector response, the atmospheric
transmission, wavelength-dependent slit losses due to atmospheric
refraction, etc. Furthermore, even with perfect calibration there are
astrophysical reasons for mismatches between models and data, such as
reddening due to dust in our Galaxy and in the source. For these
reasons we continuum filter the data by multiplying the ratio of the
data and the model by a polynomial of order
$n\equiv (\lambda_{\rm max}- \lambda_{\rm min})/100$\AA, as described
in Section 3.1.  This approach relies on the assumption that the
continuum mismatches are on scales $\gtrsim100$\AA. The index approach
relies on a similar assumption, namely that any unmodeled continuum
variation can be approximated by linear functions across the feature
of interest.  

\begin{figure*}[!t]
\center
\includegraphics[width=0.9\textwidth]{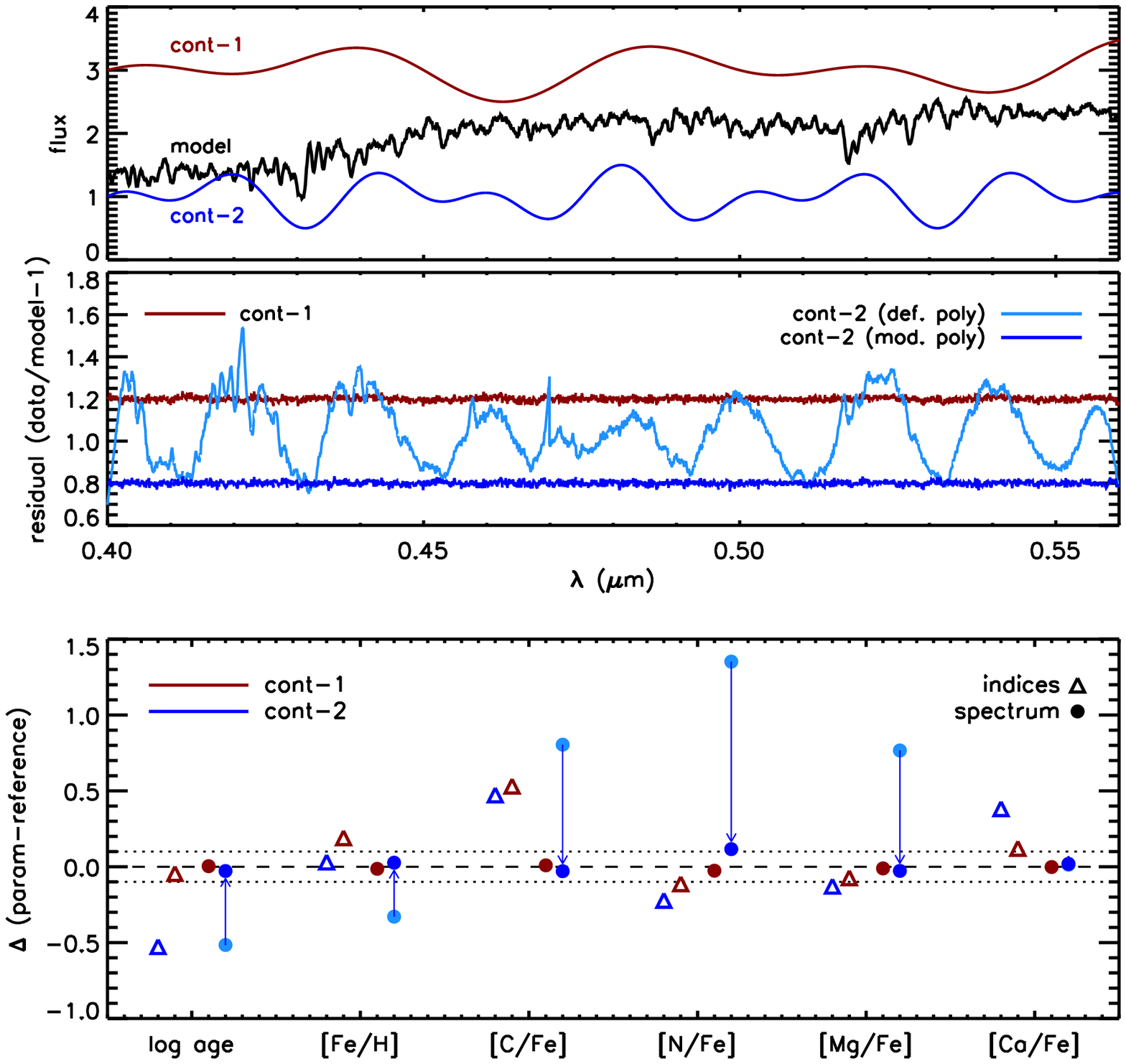}
\vspace{0.1cm}
\caption{Impact of imperfect fluxing on derived parameters.  {\it Top
    panel:} The mock spectrum (black line) and two continuum functions
  (labeled `cont-1' and `cont-2' and defined by Eqn. B1 and B2) are
  shown.  The mock spectrum is multiplied by the continuum functions
  to simulate the effect of imperfect fluxing.  {\it Middle panel:}
  Residuals between the best-fit model and the mock data for three
  cases.  The red and light blue lines show results for the default
  \alf\, setup applied to the mock data with the continuum
  modification (red is offset by +0.2).  The dark blue line shows the
  result when the polynomial model for continuum filtering in \alf\,
  is modified to enable higher order polynomials (offset by -0.2).
  {\it Bottom panel:} Best-fit parameters comparing index-based and
  full spectrum-based fitting (diamonds and circle symbols), and
  comparing the two continuum functions (red and blue lines).
  Parameters are plotted as differences with respect to parameters
  derived when no continuum function is applied to the mock spectrum.
  Points connected with arrows show results for the `cont-2' function
  with the default \alf\, polynomial model (light blue) and modified
  polynomial model (dark blue).}
\label{fig:polytest}
\end{figure*}

With the important exception of poorly modeled sky emission lines and
telluric absorption lines, there is no expectation for variations on
smaller scales.  Nevertheless, it is important to test the effects of
such variations on the measured elemental abundances.  We have
generated two continuum functions (labeled `cont-1' and `cont-2') that
are then multiplied by a mock dataset.  The resulting modified mock
spectra are then fit in \alf\, with the same approach as described in
the main text.  The two functions are defined by:
\begin{equation}
1+0.5\,{\rm sin}(2\pi\lambda/2000{\rm \AA}+e)\, {\rm sin} (2\pi\lambda/500{\rm \AA}),
\end{equation}
\noindent
and
\begin{equation}
1+0.5\, {\rm sin} (2\pi\lambda/1000{\rm \AA}+e)\, {\rm sin} (2\pi\lambda/250{\rm \AA}).
\end{equation}
\noindent 
The second function is the same as the first except with both periods
halved.  The phase of the first term was chosen in an ad hoc
manner. These functions, along with the unaltered model spectrum, are
shown in the top panel of Figure \ref{fig:polytest}.  We emphasize
that the `cont-2' model is quite extreme and pathological - in
essentially all cases of interest the effect on the continuum due to
e.g., the instrument or atmosphere is a much lower order function of
wavelength.

The middle panel of Figure \ref{fig:polytest} shows the residuals
between the mock data and the best-fit model for several cases.  The
red line shows the result of fitting the mock spectrum that was
modified by the `cont-1' function.  The light blue line shows the case
for `cont-2'.  It is immediately obvious that our default continuum
filtering approach performs well for the former and fails for the
latter case.  This is not surprising, because in \alf\, the order of the
polynomial is determined by
$n\equiv (\lambda_{\rm max}- \lambda_{\rm min})/100$\AA, which implies
approximate linearity on $\approx100$\AA\, scales.  Notice however
that the `cont-2' model is non-linear even on $\approx$100\AA\,
scales, which explains why the default \alf\, model fails to fit this
mock spectrum.

The ability to inspect the residuals in this way is a key feature of
full spectrum fitting, as the residuals often provide clues to where
and why the model has failed.  In this case it is clear that the input
data contains significant power on intermediate wavelength scales that
is not captured by our standard setup.  We demonstrate this by
considering a modified continuum filtering model where the polynomial
order is set by
$n_{\rm mod}\equiv (\lambda_{\rm max}- \lambda_{\rm min})/50$\AA.  The
residuals in this case are shown in the dark blue line in the middle
panel.  Now the residuals are much better behaved.  When fitting real
data the residuals are always carefully inspected so that situations
such as this can be identified and addressed.  Note that it is much
more difficult to identify imperfect fluxing when fitting indices, and
in practice one must rely on the data reduction delivering
imperfections that are at most linear on $\approx100$\AA\, scales.

The bottom panel of Figure \ref{fig:polytest} shows the best-fit
parameters for the two continuum-modified mock spectra.  We show
results from fitting both the full spectrum and the indices described
in Section \ref{s:spec_vs_indx}.  The parameters are plotted as
differences with respect to results from the unmodified mock spectrum
(modified - unmodified).  For the `cont-2' continuum modification we
include the same two results shown in the middle panel, namely the
default \alf\, model (light blue) and a modified version with a more
flexible polynomial model (dark blue).  With a sufficiently flexible
polynomial model (that can be determined from inspection of residuals)
the recovered parameters agree with those retrieved from the
unmodified mock spectrum.

The results from fitting the indices are fairly stable against the
broadband variations induced by the `cont-1' model.  This is not
surprising because the indices effectively assume that any fluxing
issues are linear on the scale of the index, which for most indices is
$\approx100$\AA.  The indices fair less well in the case of the
`cont-2' model, where the spectral variations induced are often
non-linear on $<100$\AA\, scales and therefore not well-modeled within
the index framework.  Perhaps more importantly, it would not be
obvious that the model has failed in this case, absent external
calibrating data to indicate that the fluxing is badly off.

The conclusion we draw from these tests is that the continuum
filtering applied in our fitting program is capable of describing
fairly pathological fluxing issues that impart features on $>50$\AA\,
scales.   Inspection of residuals is important for identifying the
most severe cases, and when the residuals are well-behaved the
resulting parameter estimates are likewise well-behaved.  When fluxing
problems introduce non-linear features on wavelength scales within an
individual index then the index will be biased in an unquantifiable
manner unless independent fluxing information is available.


\section{Model behavior in the NIR}

\begin{figure}[!t]
\center
\includegraphics[width=0.45\textwidth]{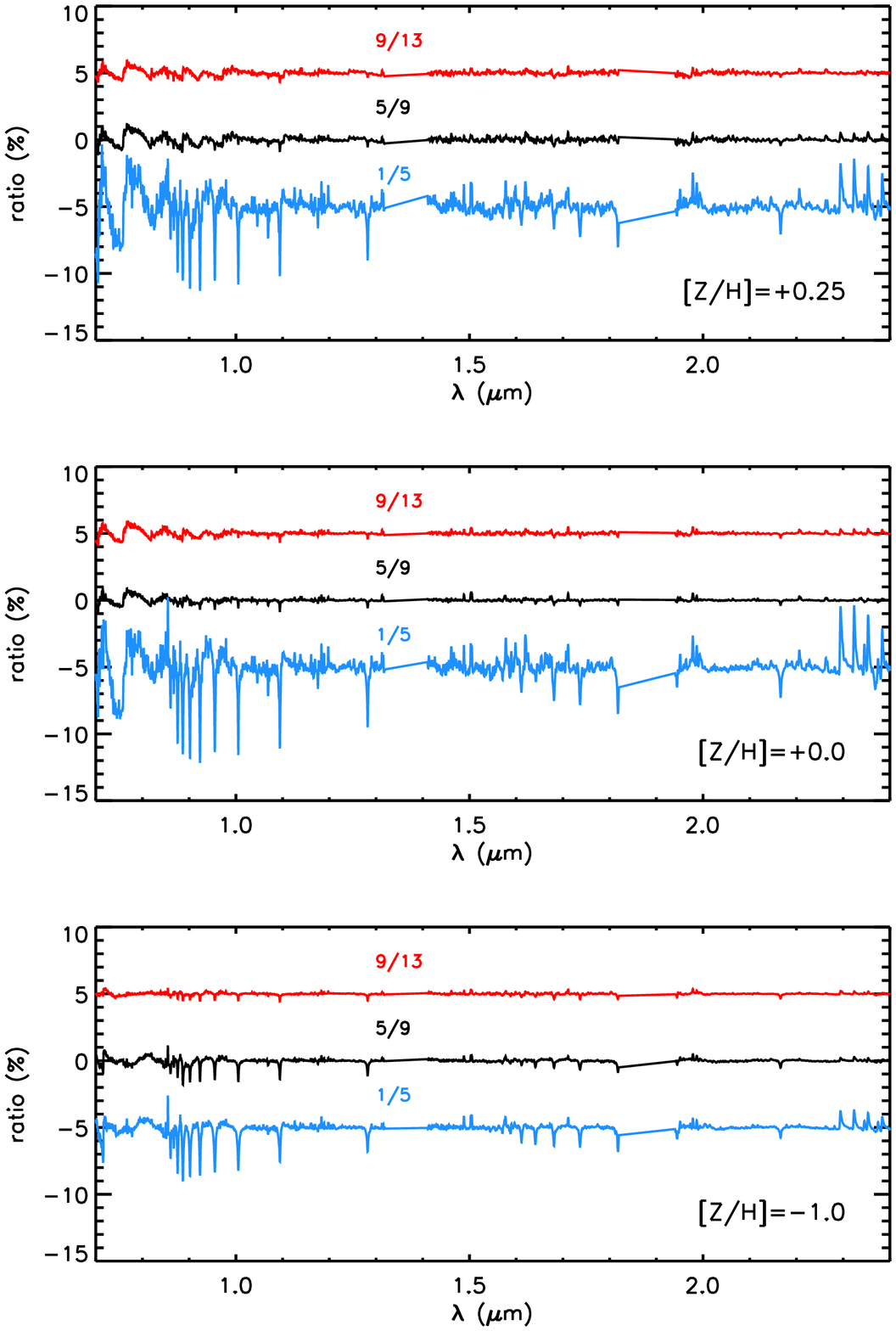}
\vspace{0.1cm}
\caption{As in Figure \ref{fig:sspage} for the $0.7-2.4\mu m$ spectral range.}
\label{fig:sspage_NIR}
\end{figure}

\begin{figure}[!t]
\center
\includegraphics[width=0.45\textwidth]{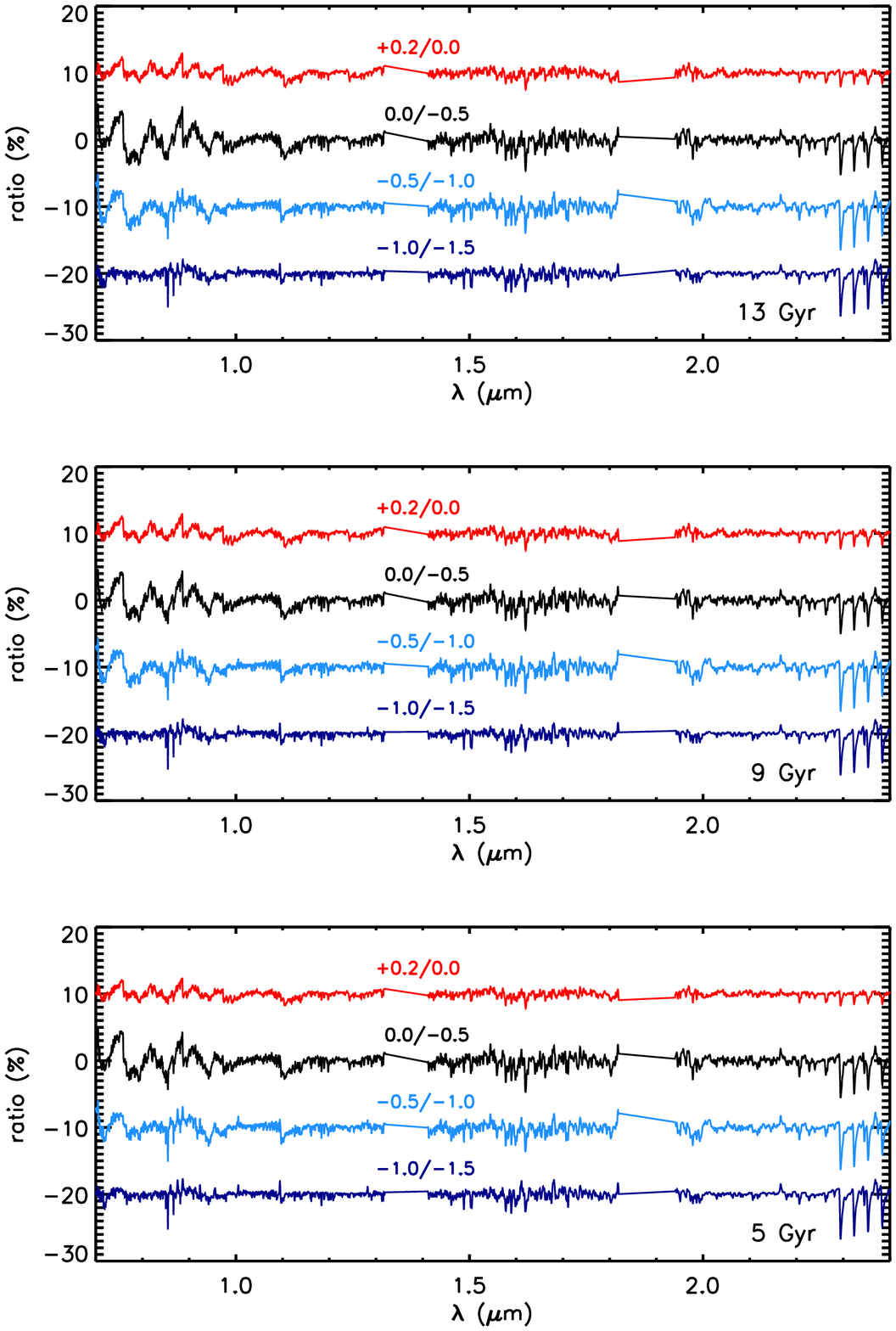}
\vspace{0.1cm}
\caption{As in Figure \ref{fig:sspzmet} for the $0.7-2.4\mu m$ spectral range.}
\label{fig:sspzmet_NIR}
\end{figure}

\begin{figure*}[t!]
\center
\includegraphics[width=0.9\textwidth]{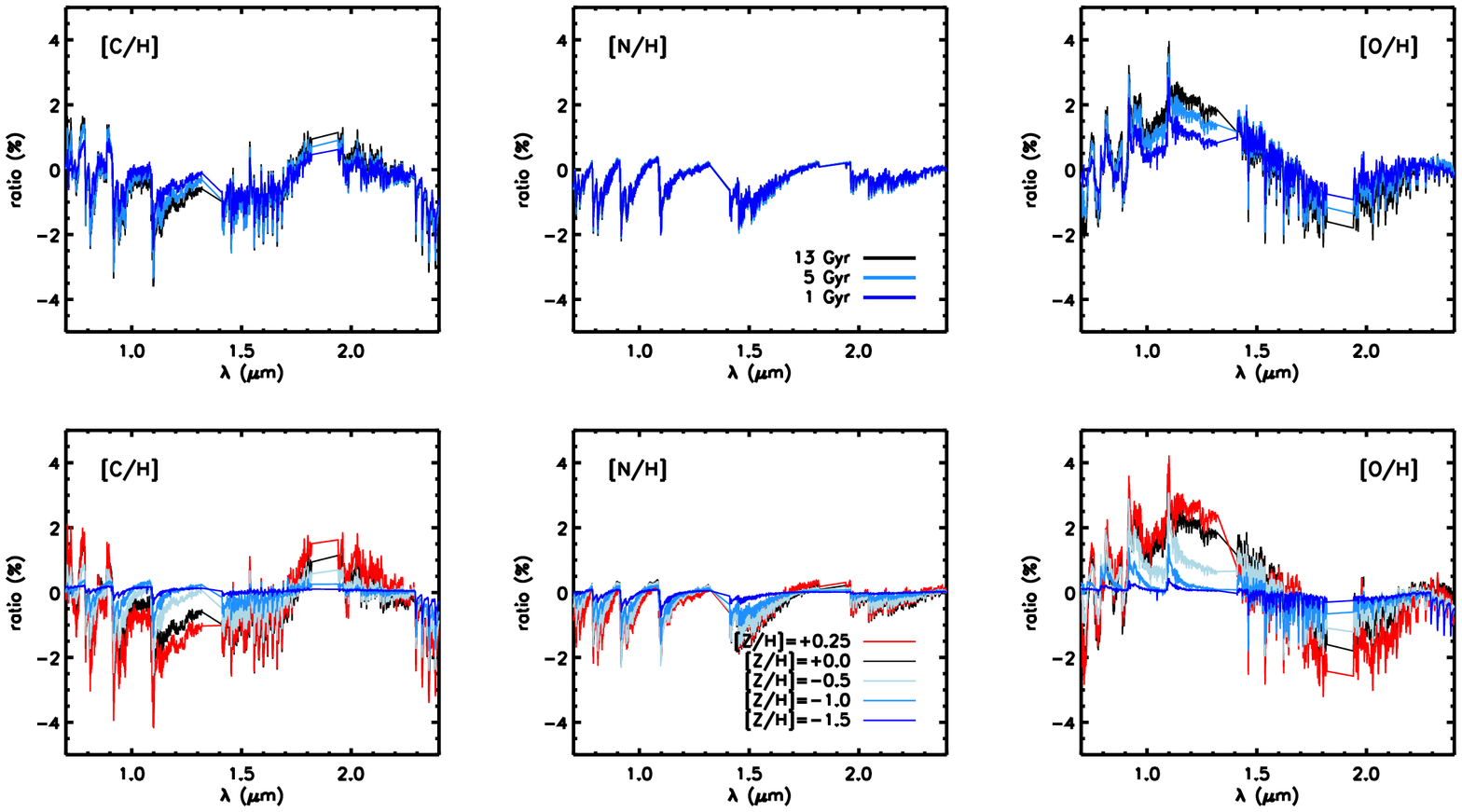}
\vspace{0.1cm}
\caption{As in Figure \ref{fig:cno} for the $0.7-2.4\mu m$ spectral range.}
\label{fig:cno_nir}
\end{figure*}

\begin{figure*}[t!]
\center
\includegraphics[width=0.9\textwidth]{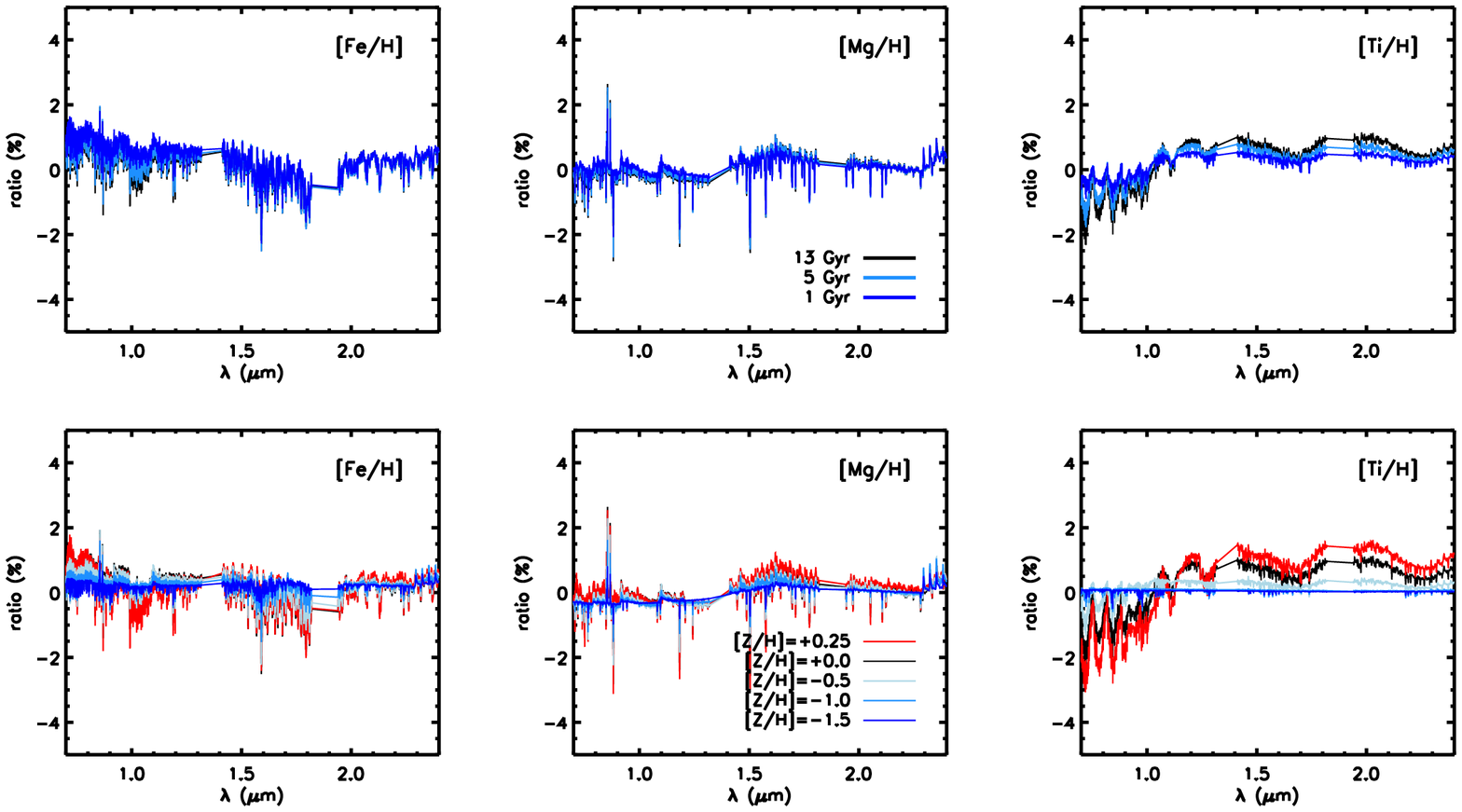}
\vspace{0.1cm}
\caption{As in Figure \ref{fig:femgti} for the $0.7-2.4\mu m$ spectral range.}
\label{fig:femgti_nir}
\end{figure*}

This appendix shows the behavior of the models in the $0.7-2.4\mu m$
region.  Figures \ref{fig:sspage_NIR} and \ref{fig:sspzmet_NIR} show
the age and metallicity sensitivity of the empirical SSPs (analogous
to Figures \ref{fig:sspage} and \ref{fig:sspzmet} in the main text).
Figures \ref{fig:cno_nir} and \ref{fig:femgti_nir} show the element
sensitivity as a function of age and metallicity for the elements C,
N, O, Fe, Mg and Ti (analogous to Figures \ref{fig:cno} and
\ref{fig:femgti} in the main text).

\end{appendix}



\end{document}